\documentclass[aps,prd,twocolumn,superscriptaddress,showpacs,preprintnumbers,amsmath,amssymb]{revtex4}
\usepackage{graphicx}
\usepackage{dcolumn}
\usepackage{bm}
\usepackage{enumerate}
\def\Pomeron{I\hspace{-1.2mm}P}
\def\met{\not\!\!E_T}
\begin{document}
\title{Diffractive dijet production 
in $\bar{p} p$ collisions at $\sqrt{s}$=1.96~TeV}
\affiliation{Institute of Physics, Academia Sinica, Taipei, Taiwan 11529, Republic of China}
\affiliation{Argonne National Laboratory, Argonne, Illinois 60439, USA}
\affiliation{University of Athens, 157 71 Athens, Greece}
\affiliation{Institut de Fisica d'Altes Energies, ICREA, Universitat Autonoma de Barcelona, E-08193, Bellaterra (Barcelona), Spain}
\affiliation{Baylor University, Waco, Texas 76798, USA}
\affiliation{Istituto Nazionale di Fisica Nucleare Bologna, $^{ee}$University of Bologna, I-40127 Bologna, Italy}
\affiliation{University of California, Davis, Davis, California 95616, USA}
\affiliation{University of California, Los Angeles, Los Angeles, California 90024, USA}
\affiliation{Instituto de Fisica de Cantabria, CSIC-University of Cantabria, 39005 Santander, Spain}
\affiliation{Carnegie Mellon University, Pittsburgh, Pennsylvania 15213, USA}
\affiliation{Enrico Fermi Institute, University of Chicago, Chicago, Illinois 60637, USA}
\affiliation{Comenius University, 842 48 Bratislava, Slovakia; Institute of Experimental Physics, 040 01 Kosice, Slovakia}
\affiliation{Joint Institute for Nuclear Research, RU-141980 Dubna, Russia}
\affiliation{Duke University, Durham, North Carolina 27708, USA}
\affiliation{Fermi National Accelerator Laboratory, Batavia, Illinois 60510, USA}
\affiliation{University of Florida, Gainesville, Florida 32611, USA}
\affiliation{Laboratori Nazionali di Frascati, Istituto Nazionale di Fisica Nucleare, I-00044 Frascati, Italy}
\affiliation{University of Geneva, CH-1211 Geneva 4, Switzerland}
\affiliation{Glasgow University, Glasgow G12 8QQ, United Kingdom}
\affiliation{Harvard University, Cambridge, Massachusetts 02138, USA}
\affiliation{Division of High Energy Physics, Department of Physics, University of Helsinki and Helsinki Institute of Physics, FIN-00014, Helsinki, Finland}
\affiliation{University of Illinois, Urbana, Illinois 61801, USA}
\affiliation{The Johns Hopkins University, Baltimore, Maryland 21218, USA}
\affiliation{Institut f\"{u}r Experimentelle Kernphysik, Karlsruhe Institute of Technology, D-76131 Karlsruhe, Germany}
\affiliation{Center for High Energy Physics: Kyungpook National University, Daegu 702-701, Korea; Seoul National University, Seoul 151-742, Korea; Sungkyunkwan University, Suwon 440-746, Korea; Korea Institute of Science and Technology Information, Daejeon 305-806, Korea; Chonnam National University, Gwangju 500-757, Korea; Chonbuk National University, Jeonju 561-756, Korea}
\affiliation{Ernest Orlando Lawrence Berkeley National Laboratory, Berkeley, California 94720, USA}
\affiliation{University of Liverpool, Liverpool L69 7ZE, United Kingdom}
\affiliation{University College London, London WC1E 6BT, United Kingdom}
\affiliation{Centro de Investigaciones Energeticas Medioambientales y Tecnologicas, E-28040 Madrid, Spain}
\affiliation{Massachusetts Institute of Technology, Cambridge, Massachusetts 02139, USA}
\affiliation{Institute of Particle Physics: McGill University, Montr\'{e}al, Qu\'{e}bec, Canada H3A~2T8; Simon Fraser University, Burnaby, British Columbia, Canada V5A~1S6; University of Toronto, Toronto, Ontario, Canada M5S~1A7; and TRIUMF, Vancouver, British Columbia, Canada V6T~2A3}
\affiliation{University of Michigan, Ann Arbor, Michigan 48109, USA}
\affiliation{Michigan State University, East Lansing, Michigan 48824, USA}
\affiliation{Institution for Theoretical and Experimental Physics, ITEP, Moscow 117259, Russia}
\affiliation{University of New Mexico, Albuquerque, New Mexico 87131, USA}
\affiliation{The Ohio State University, Columbus, Ohio 43210, USA}
\affiliation{Okayama University, Okayama 700-8530, Japan}
\affiliation{Osaka City University, Osaka 588, Japan}
\affiliation{University of Oxford, Oxford OX1 3RH, United Kingdom}
\affiliation{Istituto Nazionale di Fisica Nucleare, Sezione di Padova-Trento, $^{ff}$University of Padova, I-35131 Padova, Italy}
\affiliation{University of Pennsylvania, Philadelphia, Pennsylvania 19104, USA}
\affiliation{Istituto Nazionale di Fisica Nucleare Pisa, $^{gg}$University of Pisa, $^{hh}$University of Siena and $^{ii}$Scuola Normale Superiore, I-56127 Pisa, Italy}
\affiliation{University of Pittsburgh, Pittsburgh, Pennsylvania 15260, USA}
\affiliation{Purdue University, West Lafayette, Indiana 47907, USA}
\affiliation{University of Rochester, Rochester, New York 14627, USA}
\affiliation{The Rockefeller University, New York, New York 10065, USA}
\affiliation{Istituto Nazionale di Fisica Nucleare, Sezione di Roma 1, $^{jj}$Sapienza Universit\`{a} di Roma, I-00185 Roma, Italy}
\affiliation{Rutgers University, Piscataway, New Jersey 08855, USA}
\affiliation{Texas A\&M University, College Station, Texas 77843, USA}
\affiliation{Istituto Nazionale di Fisica Nucleare Trieste/Udine, I-34100 Trieste, $^{kk}$University of Udine, I-33100 Udine, Italy}
\affiliation{University of Tsukuba, Tsukuba, Ibaraki 305, Japan}
\affiliation{Tufts University, Medford, Massachusetts 02155, USA}
\affiliation{University of Virginia, Charlottesville, Virginia 22906, USA}
\affiliation{Waseda University, Tokyo 169, Japan}
\affiliation{Wayne State University, Detroit, Michigan 48201, USA}
\affiliation{University of Wisconsin, Madison, Wisconsin 53706, USA}
\affiliation{Yale University, New Haven, Connecticut 06520, USA}

\author{T.~Aaltonen}
\affiliation{Division of High Energy Physics, Department of Physics, University of Helsinki and Helsinki Institute of Physics, FIN-00014, Helsinki, Finland}
\author{M.~Albrow}
\affiliation{Fermi National Accelerator Laboratory, Batavia, Illinois 60510, USA}
\author{B.~\'{A}lvarez~Gonz\'{a}lez$^z$}
\affiliation{Instituto de Fisica de Cantabria, CSIC-University of Cantabria, 39005 Santander, Spain}
\author{S.~Amerio}
\affiliation{Istituto Nazionale di Fisica Nucleare, Sezione di Padova-Trento, $^{ff}$University of Padova, I-35131 Padova, Italy}
\author{D.~Amidei}
\affiliation{University of Michigan, Ann Arbor, Michigan 48109, USA}
\author{A.~Anastassov$^x$}
\affiliation{Fermi National Accelerator Laboratory, Batavia, Illinois 60510, USA}
\author{A.~Annovi}
\affiliation{Laboratori Nazionali di Frascati, Istituto Nazionale di Fisica Nucleare, I-00044 Frascati, Italy}
\author{J.~Antos}
\affiliation{Comenius University, 842 48 Bratislava, Slovakia; Institute of Experimental Physics, 040 01 Kosice, Slovakia}
\author{G.~Apollinari}
\affiliation{Fermi National Accelerator Laboratory, Batavia, Illinois 60510, USA}
\author{J.A.~Appel}
\affiliation{Fermi National Accelerator Laboratory, Batavia, Illinois 60510, USA}
% ...

\collaboration{CDF Collaboration\footnote{With visitors from
$^a$Istituto Nazionale di Fisica Nucleare, Sezione di Cagliari, 09042 Monserrato (Cagliari), Italy,
$^b$University of California Irvine, Irvine, CA 92697, USA,
$^c$University of California Santa Barbara, Santa Barbara, CA 93106, USA,
$^d$University of California Santa Cruz, Santa Cruz, CA 95064, USA,
$^e$Institute of Physics, Academy of Sciences of the Czech Republic, Czech Republic,
$^f$CERN, CH-1211 Geneva, Switzerland,
$^g$Cornell University, Ithaca, NY 14853, USA,
$^h$University of Cyprus, Nicosia CY-1678, Cyprus,
$^i$Office of Science, U.S. Department of Energy, Washington, DC 20585, USA,
$^j$University College Dublin, Dublin 4, Ireland,
$^k$ETH, 8092 Zurich, Switzerland,
$^l$University of Fukui, Fukui City, Fukui Prefecture, Japan 910-0017,
$^m$Universidad Iberoamericana, Mexico D.F., Mexico,
$^n$University of Iowa, Iowa City, IA 52242, USA,
$^o$Kinki University, Higashi-Osaka City, Japan 577-8502,
$^p$Kansas State University, Manhattan, KS 66506, USA,
$^q$Korea University, Seoul, 136-713, Korea,
$^r$University of Manchester, Manchester M13 9PL, United Kingdom,
$^s$Queen Mary, University of London, London, E1 4NS, United Kingdom,
$^t$University of Melbourne, Victoria 3010, Australia,
$^u$Muons, Inc., Batavia, IL 60510, USA,
$^v$Nagasaki Institute of Applied Science, Nagasaki, Japan,
$^w$National Research Nuclear University, Moscow, Russia,
$^x$Northwestern University, Evanston, IL 60208, USA,
$^y$University of Notre Dame, Notre Dame, IN 46556, USA,
$^z$Universidad de Oviedo, E-33007 Oviedo, Spain,
$^{aa}$CNRS-IN2P3, Paris, F-75205 France,
$^{bb}$Texas Tech University, Lubbock, TX 79609, USA,
$^{cc}$Universidad Tecnica Federico Santa Maria, 110v Valparaiso, Chile,
$^{dd}$Yarmouk University, Irbid 211-63, Jordan,
}}
\noaffiliation
\affiliation{CDF collaboration}

\date{\today}
\begin{abstract}
 We report on a study of diffractive dijet production in $\bar pp$  collisions at $\sqrt s=1.96$ TeV using the CDF~II detector at the Fermilab Tevatron $\bar pp$ collider. A data sample from 310 pb$^{-1}$ of integrated luminosity collected by triggering on a high transverse energy jet, $E_T^{jet}$,  in coincidence with a recoil antiproton detected in a Roman pot spectrometer is used to measure the ratio of single-diffractive to inclusive-dijet event rates as a function of $x^{\bar p}$ of the interacting parton in the antiproton, the Bjorken-$x$, $x^{\bar p}_{Bj}$, and a $Q^2\approx (E_T^{jet})^2$ in the ranges $10^{-3}<x^{\bar p}_{Bj}<10^{-1}$ and $10^2<Q^2 <10^4$ GeV$^2$, respectively. Results are presented for the region of $\bar p$-momentum-loss fraction $0.03<\xi_{\bar p}<0.09$ and a four-momentum transfer squared $t_{\bar p}>-4$~GeV$^2$. The $t_{\bar p}$ dependence is measured as a function of $Q^2$ and $x_{Bj}^{\bar p}$ and compared with that of inclusive single diffraction dissociation. We find weak $x^{\bar p}_{Bj}$ and $Q^2$ dependencies in the ratio of single diffractive to inclusive event rates, and no significant $Q^2$ dependence in the diffractive $t_{\bar p}$ distributions.    
\end{abstract}

\maketitle
\section{Introduction}\label{introduction}
We report on measurements performed on diffractive dijet production in proton-antiproton collisions at $\sqrt s=1.96$~TeV at the Fermilab Tevatron in Run~II using the CDF~II detector. The CDF collaboration has previously reported several results on low-transverse-momentum (soft) and high-transverse-momentum (hard) diffractive processes, obtained at the Fermilab Tevatron $\bar pp$ collider in Run~I (1992--96)~\cite{elastic}-\cite{idpe} and in Run~II (2001--06)~\cite{CDFgammagamma}-\cite{CDFgammagamma2}. Three of the Run~I results~\cite{jj,jjRPS,jjRPS630} are based on dijet events, $JJ$, produced in single diffraction (or single dissociation, SD). These events are characterized by the presence of two jets in the final state and a  leading  $\bar p$  that escapes the collision intact. A large rapidity gap, $G_{\bar p}$, defined as a region of pseudorapidity~\cite{rapidity} devoid of particles, is present between the surviving $\bar p$  and the diffractive cluster, $X_p$, composed of  the particles from the dissociation of the proton, including $JJ$:

\begin{equation}
\bar p+p\rightarrow \bar p+G_{\bar p}+X_p, \:\:X_p\rightarrow X+JJ,
\end{equation}

\noindent where $X$ represents all other particles in the diffractive cluster and will be referred below as the underlying event (UE).   

Among the results of the Run~I diffractive measurements, the most striking one is the observation of a breakdown of QCD factorization, expressed as a suppression by a factor of ${\cal{O}}(10)$ of the diffractive structure function (DSF) measured in dijet production relative to that derived from fits to parton  densities measured in diffractive deep inelastic scattering (DDIS) at the DESY $e$-$p$ collider HERA (see~\cite{jjRPS}). This result was further explored by CDF in Run~I in studies of other diffractive processes, as briefly discussed below.  

In this paper, we study diffractive dijet production in Run~II with an upgraded CDF detector, incorporating special forward detector-components that expand the kinematic reach of the variables that define the process. Our goal is to  further  characterize the properties of diffractive dijet production in an effort to decipher the QCD nature of the diffractive exchange, traditionally referred to as Pomeron ($\Pomeron$) exchange~\cite{Collins}-\cite{Donnachie}. Below, we first present a brief physics-oriented summary of the diffractive processes studied and the results obtained in Run~I, which is intended to define the terms and concepts used in the present study. Then, we restate the aim of this study in light of the Run~I results, and conclude with an outline of the organization of the paper.

In Run~I, several soft and hard diffraction processes were studied at $\sqrt s=1800$~GeV,
and in some cases at $\sqrt s =630$ GeV. 
Two types of hard diffraction results were obtained:

\newcounter{Lcount}
\begin{list}{(\alph{Lcount})}
{\usecounter{Lcount}
 \setlength{\rightmargin}{\leftmargin}}
\addtolength{\itemsep}{-0.5em}
\item gap results: diffractive to nondiffractive (ND) cross
section ratios using the (large) rapidity gap signature
to select diffractive events;
\item RPS results: diffractive to ND structure function ratios using a
Roman Pot Spectrometer (RPS) to trigger on and measure a leading antiproton. 
\end{list}

\begin{figure}[h]
\begin{center}
\includegraphics[width=0.5\textwidth,bb=41 313 467 445]{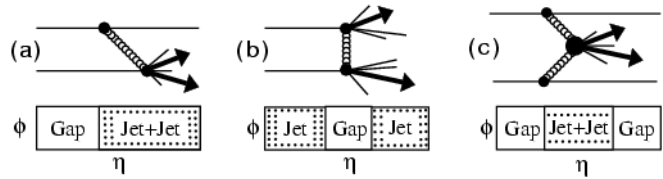}
\caption{\label{fig:diagrams} 
Leading order schematic diagrams and event topologies in pseudorapidity ($\eta$) vs azimuthal angle ($\phi)$ for diffractive dijet production processes studied by CDF: (a) single diffraction, (b) double diffraction, and  (c) double Pomeron exchange; the dot-filled  rectangles represent regions where particle production occurs.}
\end{center}
\end{figure}

The diffractive dijet production processes studied are shown schematically in Fig.~\ref{fig:diagrams}. They include (a) single diffraction (or single dissociation, SD), (b) double diffraction (or double dissociation, DD), and (c) double Pomeron exchange (DPE) or central dissociation (CD). For SD, results have also been obtained for  $W$~\cite{W}, $b$-quark~\cite{b-quark}, and  $J/\psi$~\cite{jpsi} production.

Soft diffraction processes studied include SD~\cite{sd}, DD~\cite{dd}, DPE (or CD)~\cite{idpe}, and single plus double diffraction (SDD)~\cite{sdd}. These processes are defined as follows:

\begin{tabular}{ll}
{\bf SD}&$\bar{p}p\rightarrow \bar{p}+G_{\bar p}+X_p$,\\
{\bf DD}&$\bar{p}p\rightarrow X_{\bar p}+G_c+X_p,$\\
{\bf DPE} or {\bf CD}&$\bar{p}p\rightarrow \bar{p}+G_{\bar p}+X_c+G_p+p$,\\
{\bf SDD}&$\bar{p}p\rightarrow \bar{p}+G_{\bar p}+X_c+G_{X_p}+X_p$,
\end{tabular}
 
\noindent where $X_{\bar p}$, $X_p$ and $X_c$ are antiproton-, proton- and central-dissociation particle clusters, $G_c$ is a central gap, and $G_{X_p}$ is a gap adjacent to a proton-dissociation cluster.

The results obtained exhibit regularities in normalization
and factorization properties that point to
the QCD character of diffraction~\cite{dino}. For example, 
at $\sqrt s=$1800 GeV the SD/ND ratios, also referred to as diffractive- or gap-fractions,   
for dijet, $W$, $b$-quark, and $J/\psi$ production, 
as well as the ratio of DD/ND dijet-production, 
are all $\approx 1\%$. This value is suppressed relative 
to standard QCD-inspired theoretical 
expectations ({\em e.g.,} two-gluon exchange)
by a factor of  ${\cal{O}}(10)$, which is comparable to the suppression factor observed in soft diffraction relative to Regge-theory-based predictions.
However, except for an overall suppression in normalization, factorization approximately holds among various processes at fixed $\sqrt s$~\cite{dino}.

In this paper, we report on the dependence on Bjorken-$x$, $Q^2$ and  $t$, the square of the four-momentum transfer between the incoming and outgoing antiproton, of the diffractive structure function measured from dijet production in association with a leading antiproton in $\bar{p} p$ collisions at $\sqrt s$=1.96 TeV and compare our results with those from HERA and with theoretical expectations. 

The paper is organized as follows: in Sec.~\ref{method} we discuss the method we use to extract the diffractive structure function; in Sec.~\ref{apparatus} we describe the experimental apparatus; in Secs.~\ref{data} and~\ref{analysis} we discuss the data sets and data analysis; in Sec.~\ref{results} we present the results; and in Sec.~\ref{conclusion} we summarize the results and draw conclusions. 

\section{Method\label{method}}
The cross section for inclusive dijet production in $\bar p p$ collisions can be written as 

\begin{equation}\label{eq:nd}
\frac{d^3\sigma_{jj}^{\rm incl}}{dx_{\bar p} dx_p d\hat{t}} = \frac{F^{\rm incl}_{jj}(x_{\bar p},Q^2)}{x_{\bar p}} \cdot
\frac{F^{\rm incl}_{jj}(x_{p},Q^2)}{x_{p}} \cdot
\frac{
d\hat\sigma_{jj}
}
{d\hat{t}},
\end{equation}

\noindent
where $x_{\bar p}$ ($x_{p}$) is the Bjorken-$x$, defined as the forward-momentum fraction of the interacting parton of the antiproton (proton), $Q^2$ is the factorization hard scale, $F^{\rm incl}_{jj}(x_{\bar p},Q^2)$ and $F^{\rm incl}_{jj}(x_{p},Q^2)$ are structure functions, and $\hat\sigma_{jj}$ is the scattering cross section of the two partons producing the final-state jets. The $\hat\sigma_{jj}$ depends on $x_{\bar p}$ and $x_p$, and also on $\hat{t}$, the square of the four-momentum transfer between the interacting partons, on $Q^2$, and on $\alpha_s(Q^2$), the strong-interaction running coupling constant.  

The structure function relevant for dijet production is a color-weighted combination of gluon ($g$) and quark ($q$) terms given by:
 
\begin{equation}
F^{\rm incl}_{jj}(x, Q^2)= x \left[g(x, Q^2)+ \frac{4}{9} \sum_i q_i(x, Q^2)\right],
\end{equation}
where $x$ refers to $x_p$ or $x_{\bar p}$.
In analogy with Eq.~(\ref{eq:nd}), the differential cross section for diffractive dijet production can be written as~\cite{dino}

\begin{equation}
\frac{d^5\sigma^{\rm SD}_{jj}}
{dx_{\bar p} dx_{p} d\hat{t}d\xi dt} =
\frac{F_{jj}^{\rm SD} (x_{\bar p}, Q^2, \xi, t)}
{x_{\bar p}}\cdot\frac{F^{\rm incl}_{jj} (x_{p}, Q^2)}{x_{p}} \cdot
\frac{d\hat{\sigma}_{jj}
}{d\hat{t}},
\end{equation}

\noindent
where $F_{jj}^{\rm SD}$ is the diffractive structure function, which in addition to the usual dependence on $x_{\bar p}$ and $Q^2$ also depends on $\xi$, the fractional forward momentum loss of the antiproton, and on $t$. 

The jet energies measured in the CDF detector must be corrected for various detector effects, which depend on the jet energy and $\eta$-$\phi$ coordinates due to differences in calorimeter subsystem designs and calorimeter interfaces~\cite{jetcorrections}. The correction factor generally increases as the jet energy decreases. To avoid systematic uncertainties associated with estimating corrections using Monte Carlo simulations, particularly for diffractively produced jets of relatively low $E_T$, we measure ratios of SD to inclusive production as a function of Bjorken-$x$ and $Q^2$, for which jet-energy corrections due to  detector effects nearly cancel out. 

Since the single-diffractive cross section is a small fraction of the inclusive one, $\lessapprox 1$\% of $\sigma^{\rm incl}$, we refer below to ``inclusive'' and ``nondiffractive'' (ND) dijet-production interchangeably. Furthermore, we also use this approximation in classifying event samples and associated parameters, as for example between $F_{jj}^{\rm incl}$ and $F_{jj}^{\rm ND}$. 

The diffractive structure function is obtained by multiplying 
the ratio ${\rm R}_{\rm {\rm SD/ND}}(x, \xi, t)$ of the SD to ND event densities, 
$n_{jj}^{\rm SD}(x, Q^2, \xi, t)$ and $n^{\rm ND}_{jj}(x, Q^2)$, by the ND structure function, $F^{\rm ND}_{jj}(x, Q^2)$:

\begin{equation}
F_{jj}^{\rm SD}(x, Q^2, \xi, t)={\rm R}_{\rm {SD}/{ND}}(x, \xi, t)\times F^{\rm ND}_{jj}(x, Q^2).
\label{eq:RjjSD}
\end{equation}

\noindent
This method of measuring diffractive structure functions relies on the LO QCD expectation that cross sections are proportional to structure functions:

\begin{equation}
{\rm R}_{\rm {SD}/{ND}}(x, Q^2, \xi, t)= \frac{n_{jj}^{\rm SD}(x, Q^2, \xi, t)}{n^{ND}_{jj}(x, Q^2)}
\approx \frac{F_{jj}^{\rm SD}(x, Q^2, \xi, t)}{F^{\rm ND}_{jj}(x, Q^2)}.
\label{eq:Rjj}
\end{equation}

 Next-to-leading-order corrections to ${F_{jj}^{\rm SD}(x, Q^2, \xi, t)}$ obtained by this method are expected to be of $\cal{O}$(10\%)~\cite{ken_thesis}. 

\section{Experimental apparatus\label{apparatus}}

The CDF~II detector is equipped with special forward detectors~\cite{excl2j,MP_prototype_test,forward_proposal,miniplug_nim,michele_rio} designed to enhance the capabilities for studies of diffractive physics. These detectors 
include the RPS, the Beam Shower Counters (BSC), and the MiniPlug (MP) calorimeters. The RPS is a scintillator fiber tracker used to detect leading antiprotons; the BSC are scintillator counters 
installed around the beam-pipe at three (four) locations along the $p$ ($\bar p$) direction and are used to identify rapidity gaps in the region 5.5~$<|\eta |<$~7.5; and the MP calorimeters~\cite{miniplug_nim} are two lead-scintillator based forward calorimeters covering the pseudorapidity region 3.6~$<|\eta |<$~5.1. 
The forward detectors include a system of Cherenkov Luminosity Counters (CLC)~\cite{CLC}, whose primary function is to measure the number of inelastic $\bar{p} p$ collisions per beam-bunch crossing and thereby the luminosity. 
 The CLC covers the range $3.7<|\eta|<4.7$, which substantially overlaps the MP coverage. In this analysis, the CLC is used for diagnostic purposes, and also to refine the rapidity-gap definition by detecting charged particles that might penetrate a MP without interacting, yielding a pulse-height smaller than the MP tower thresholds. 

\begin{figure*}
\begin{center}
\includegraphics[width=1\textwidth]{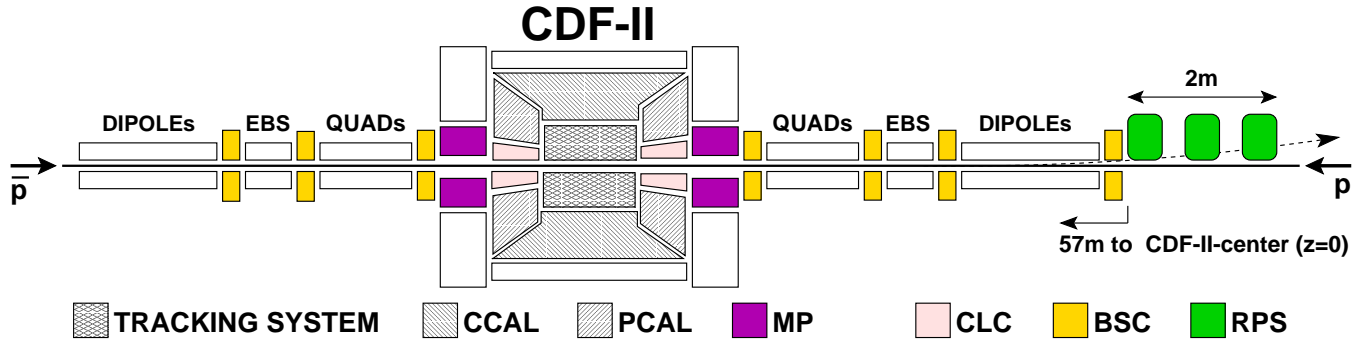}
\caption{\label{fig:fd_top} 
Schematic plan view of the CDF~II central (CDF-II) and forward detectors (acronyms are defined in the text).}
\end{center}
\end{figure*}

Figure~\ref{fig:fd_top} 
shows a schematic plan view of the beamline elements and forward detectors along the outgoing antiproton beam direction. 
The RPS comprises three Roman pot stations located at a distance of $\sim 57$~m from the nominal interaction point (IP) along the outgoing $\bar p$ direction. Each station is equipped with one Roman Pot Trigger (RPT) counter and an 64-channel scintillator-fiber tracker, which can be used to reconstruct tracks both in the $X$ (horizontal) and $Y$ (vertical) coordinates. The tracking information, in conjunction with the interaction point coordinates, provided by the central detector, can be used to calculate the variables $\xi$ and $t$ of the recoil antiproton. During running, the RPS detectors were brought up to a distance of $X\approx 1$~cm from the outgoing antiproton beam, with the (more intense) incoming proton beam  positioned $\sim 2$~mm farther away.  

In Run~I, in which data were collected at lower luminosities, the RPS detectors were positioned at approximately the same distance from the antiproton beam, but the more intense proton beam was in between the antiproton beam and the 
RPS detectors. Background from beam-halo particles from the proton beam was limiting the RPS acceptance at low $t$. Reversing the polarity of the electrostatic beam separators (EBS) in Run~II, brought the antiproton beam in between the proton beam and the RPS. This enabled data taking at the higher instantaneous luminosities of Run~II at $\sqrt s=1.96$ TeV with approximately the same $t$ acceptance as in Run~I at $1.8$~TeV.    

The CDF-I and CDF-II main detectors are multipurpose detectors described in detail in Refs.~\cite{CDF-I,CDF-II}. The most relevant components of CDF~II for this analysis are the charged particle tracking system and the central and plug calorimeters.  The tracking system consists of a silicon vertex detector surrounded by the central outer tracker (COT)~\cite{COT}. The part of the silicon vertex detecor used is the SVX~II~\cite{SVX}, composed of double-sided microstrip silicon sensors arranged in five cylindrical shells of radii
between 2.5 and 10.6 cm. The COT is an open-cell drift chamber consisting of 96 layers organized in 8 superlayers
with alternating structures of axial and $\pm2^{\circ}$ stereo readout within a  radial range between
40 and 137 cm. Surrounding the tracking detectors is a superconducting solenoid, which provides a
magnetic field of 1.4~T. Calorimeters located outside the solenoid are physically divided into a 
central calorimeter (CCAL)~\cite{CEM, CHA}, covering the pseudorapidity range $|\eta|<1.1$, and a plug calorimeter 
(PCAL)~\cite{PCAL}, covering the region $1.1<|\eta|<3.6$. These calorimeters are segmented into projective towers pointing back to the central interaction point, with granularity $\Delta\eta\times\Delta\phi \approx 0.1\times15^{\circ}$.


\section{Data Samples and Event Selection\label{data}}
The Run~I diffractive dijet results were obtained from data samples collected at low instantaneous luminosities to minimize background from pileup events. Constrained by statistics, the $Q^2$ dependence of the DSF was measured over a limited  range. Moreover, due to uncertainties in the beam position at the RPS location, it was difficult to reliably extract normalized $t$-distributions. In Run~II, our goal was to obtain high statistics diffractive data samples with low pileup backgrounds from which to extract the $Q^2$ and $t$ dependence over a wide range. To achieve this goal, we built special forward detectors, described above in Sec.~\ref{apparatus}, implemented dedicated triggers, and developed analysis techniques for background subtraction and RPS alignment. In this Section, we present the data samples and event selection requirements applied to produce the data sets from which the results were extracted.

This analysis is based on data from an integrated luminosity $L\approx 310$~pb$^{-1}$ collected in 2002--2003.  
Only ``good runs'' with a minimum $L\approx 10$~nb$^{-1}$ are used, selected based on beam conditions, detector performance criteria, and the requirement that the following detector components be functional:  CCAL, PCAL, MP, CLC, and BSC.
On-line event selection is accomplished with a three-level trigger system (L1, L2, L3) which accepts soft-interaction events as well as hard interaction events containing high $E_T$ jets. The latter are selected at the trigger level by requiring at least one calorimeter tower with $E_T>5,\,20,\mbox{ or }50$~GeV within  $|\eta|<3.5$. Leading antiprotons with fractional momentum-loss in the range $0.03\lesssim \xi_{\overline{p}} \lesssim 0.09$ were triggered on using a three-fold coincidence of the RPT counters.

Jets were reconstructed using a cone algorithm, in which the transverse energy of a jet is defined as $E_T^{jet}\equiv \Sigma_iE_i\sin(\theta_i)$ with the sum carried over all calorimeter towers at polar angles $\theta_i$ within the jet cone. The midpoint algorithm~\cite{MidPoint} was used, which is an improved iterative cone clustering algorithm based on calorimeter towers with $E_T>100$ MeV. 
The jets were corrected for detector effects and for contributions from the underlying event (UE)~\cite{jetcorrections}. Dijet candidate events were required to  have at least two jets with $E_T>5,\,20,\mbox{ or }50$~GeV depending on the event sample, and $|\eta|<2.5$.

In order to explore the region of large-transverse-energy jets of relatively low cross sections, data samples of RPS-triggered events in conjunction with the presence of a jet with $E_T^{jet}\geq$5, 20, or 50~GeV in CCAL or PCAL were also studied. These samples are referred to as RPS$\cdot$Jet5, RPS$\cdot$Jet20, and RPS$\cdot$Jet50. Corresponding ND samples (Jet5, Jet20, and Jet50) were used for comparison. 

The majority of the data used in this analysis were recorded without RPS tracking information. For these data, 
the value of $\xi$ was evaluated from calorimeter information and will be referred to as $\xi^{CAL}_{\bar p}$ (see Sec.~\ref{sec:xisec}). The $\xi^{CAL}_{\bar p}$ was then calibrated against $\xi$ obtained from the RPS, $\xi^{RPS}_{\bar p}$,  using data from runs in which RPS tracking was available (see Sec.~\ref{sec:xical}). Below, we list the definitions of the triggers used in data acquisition and the data selection requirements applied in obtaining the data samples for this analysis.

The following trigger definitions are used:
\begin{itemize}
\renewcommand{\labelitemi}{$\bullet$}
\item RPS: triple coincidence among the three RPS trigger counters in time with a $\bar{p}$ gate; 
\item RPS$_{\rm track}$: RPS with RPS tracking available (included in the RPS trigger); 
\item J5 (J20, J50): jet with $E_T^{jet}\geq 5$ (20, 50) GeV in CCAL or PCAL;
\item RPS$\cdot$Jet5 (Jet20, Jet50): RPS in coincidence with J5 (J20, J50).
\end{itemize}

The data selection requirements are listed below:

\begin{enumerate}[(i)]
\item good-run events: accepts  events from runs with no problems caused by hardware or software failures during data acquisition;
\item $E\!\!\!\!/_T$ significance~\cite{missingET}: selects events with missing transverse energy significance $S_{E\!\!\!\!/_T}<2$ to reject jet events in which there is $E\!\!\!\!/_T$ due to energy loss in calorimeter cracks and/or events with jets and (undetected) neutrinos, such as from $W\rightarrow l\nu+jets$;
\item $N(jet)\ge 2$: accepts events with two jets of $E^{jet}_T>5$ GeV within  $|\eta^{jet}|<2.5$;
\item splash veto: rejects events that cause splashes (large number of hits) in the RPT counters;
\item RPT: rejects events that are triggered by accidental (not due to the traversal of a single particle) RPS counter coincidences (less than 0.1\%);
\item SD ($0.03 < \xi^{\rm CAL}_{\overline{p}} < 0.09$): accepts SD events with good efficiency, while rejecting backgrounds from pileup events consisting of a soft SD event that triggers the RPS and a ND dijet event. 
\end{enumerate}

\noindent Table~\ref{tab:SD_cuts} lists the number of events surviving these requirements when applied successively to the data. 

\begin{table*}[htp]
\label{table1}
\begin{center}
\caption{\label{tab:SD_cuts}The number of events surviving successive selection requirements by data sample.}
\vspace*{0.5em}
\begin{ruledtabular}
\begin{tabular}{lrrrr}
Selection requirement						& RPS 	    & RPS$\cdot$Jet5     & RPS$\cdot$Jet20    & RPS$\cdot$Jet50 \\
\hline
Trigger level					& 1\mbox{ }634\mbox{ }723 &1\mbox{ }124\mbox{ }243  & 1\mbox{ }693\mbox{ }644 & 757\mbox{ }731\\
Good-run events						& 1\mbox{ }431\mbox{ }460 &  955\mbox{ }006  & 1\mbox{ }421\mbox{ }350 & 561\mbox{ }878\\ 
$E\!\!\!\!/_T$ significance: $S_{\not E}\equiv{\not\!E_T}/\sqrt{\Sigma{E_T^2}}<2$		& 1\mbox{ }431\mbox{ }253 &  950\mbox{ }776  & 1\mbox{ }410\mbox{ }780 & 539\mbox{ }957\\
$N(jet)\ge 2$: $E^{1,2}_T>5$\mbox{ }GeV, $|\eta^{1,2}|<2.5$ 	&    59\mbox{ }157 &  557\mbox{ }615  & 1\mbox{ }168\mbox{ }881 & 521\mbox{ }645\\
Splash veto	&    27\mbox{ }686 &  259\mbox{ }186  &   541\mbox{ }031 & 215\mbox{ }975\\
RPT					&    27\mbox{ }680 &  259\mbox{ }169  &   541\mbox{ }003 & 215\mbox{ }974\\
SD ($0.03 < \xi^{CAL}_{\overline{p}} < 0.09$)			&     1\mbox{ }458 &   20\mbox{ }602  &    26\mbox{ }559 &   4\mbox{ }432\\
\end{tabular}
\end{ruledtabular}
\end{center}
\end{table*}

\begin{figure}[h]
\begin{center}
\includegraphics[width=0.5\textwidth]{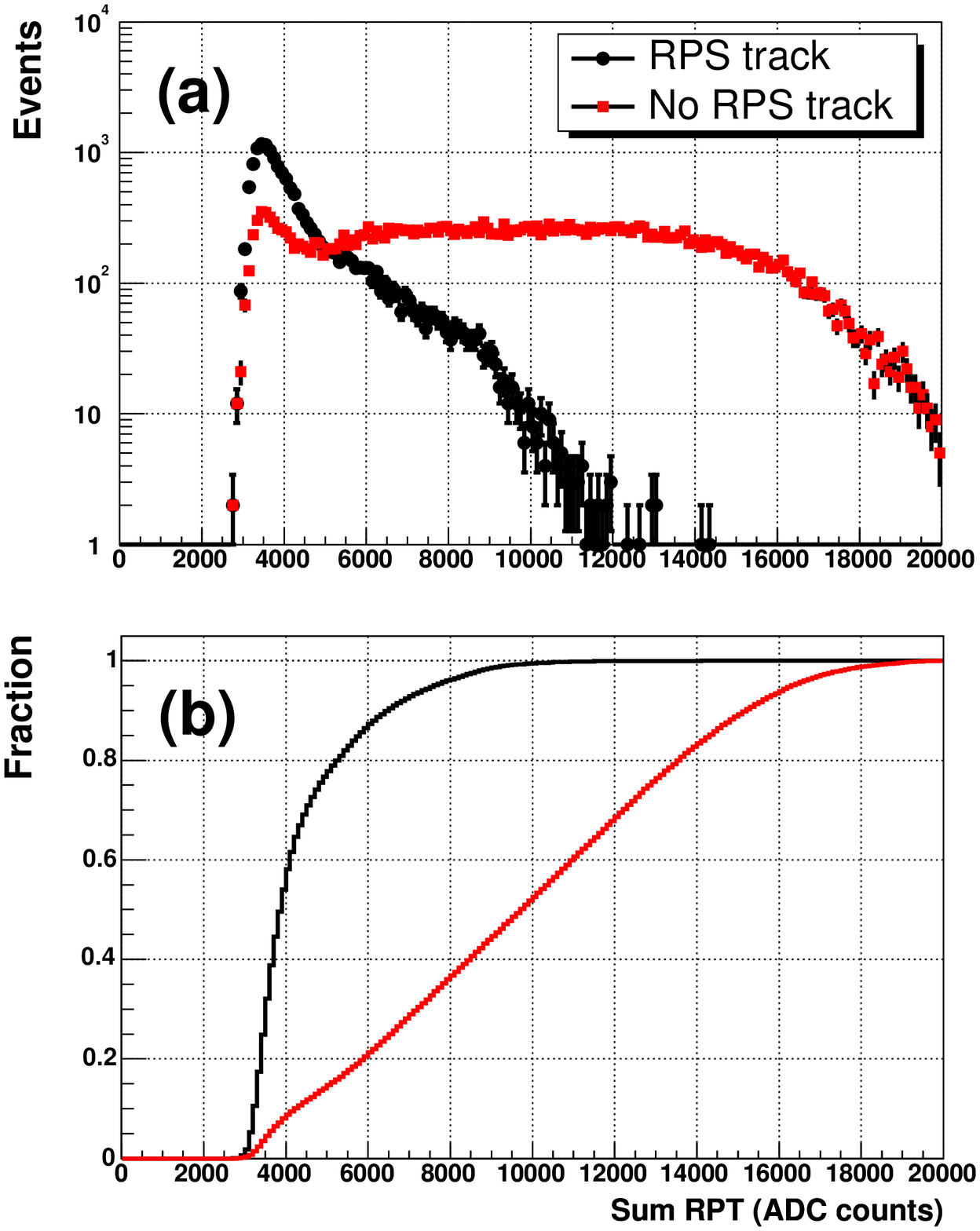}
\caption{\label{fig:sumrpt} 
SumRPT distributions for events of the SD RPS$_{\rm track}$ data sample.
(a) Number of events and (b) fraction of events as a function of the sum of ADC counts of the three Roman Pot Trigger (RPT) counters, SumRPT. In (a), the peak at SumRPT$\sim 3000$, attributed to a minimum ionizing particle (MIP) signal, is prominent in the ``RPS track'' entries, where two-MIP (three-MIP) shoulders at $\sim 6000~(9000)$ ADC counts are also discernible.  The ``No RPS track''  entries show a smaller MIP peak and a broad plateau, which is attributed to ``splash events'' (see text).}
\end{center}
\end{figure}

The splash events were studied using the RPS$_{\rm track}$ data sample before applying the splash veto requirement.  
The sum of the ADC counts of the three RPT counters, ${\rm SumRPT}\equiv\Sigma_{i=1}^3{\rm RPT}_i^{\rm ADC}$, is used to reject splash events.
Figure~\ref{fig:sumrpt}(a) shows SumRPT for events with/without a reconstructed RPS track. In the ``RPS~track'' histogram the peak at $\sim 3000$ ADC counts is attributed to a  single minimum-ionizing particle (MIP) traversing all three RPT counters. 2-MIP and 3-MIP shoulders are also discernible at $\sim 6000$ and $\sim 9000$ ADC counts, respectively. 
Events labeled as ``No~RPS~track'' are the splash events. These events dominate the region of $\rm {SumRPT}>5000$.
Detailed studies using various event samples indicate that splash events are likely to be due
to high-$\xi_{\bar p}$ diffractive events for which the $\bar p$ does not reach the RPS, but rather   
interacts with the material of the beam pipe in the vicinity of the RPS producing a spray of particles causing the splash.
In the region of ${\rm SumRPT}<5000$, the ``No~RPS~track'' distribution
has a peak similar to that seen in the  events with a reconstructed RPS track. 
These events are interpreted as good events for which the track was not reconstructed due to either malfunction or inefficiency of the fiber tracker. 

Events with SumRPT$>5000$ are rejected. 
The retained events contain approximately 77\% of the RPT sample, as can be qualitatively seen in Fig.~\ref{fig:sumrpt}(b). These events, after applying the RPT selection requirement, constitute the SD event samples listed in Table~\ref{tab:SD_cuts}.
Any possible inefficiency caused by the SumRPT cut is taken into account by folding a 6\% uncertainty into that of the extracted cross section (see Sec.~\ref{xbj_dependence}, Table~\ref{tab:systematics}).

\section{Data analysis\label{analysis}}
Analysis details are presented in seven subsections organized by measurement topic, as follows:

\begin{list}{\Alph{Lcount} --}
{\usecounter{Lcount}
 \setlength{\rightmargin}{\leftmargin}}
\addtolength{\itemsep}{-0.75em}
\vspace*{-0.5em}
\item Dynamic alignment of Roman pos spectrometer
\item Trigger efficiency of Roman pot spectrometer
\item Antiproton momentum-loss measurement
\item Multiple interactions
\item Calibration of $\xi_{\bar p}^{\rm CAL}$
\item MiniPlug contribution to $\xi_{\bar p}^{\rm CAL}$
\item Beam-halo background
\end{list}

\subsection{Dynamic alignment of Roman pot spectrometer\label{sec:align}}
The values of both  $\xi$ and $t$ can be accurately determined from RPS reconstructed track coordinates and the position of the event vertex at the Interaction Point using the beam-transport matrix between the Interaction Point and RPS. Crucial for this determination is the detector X--Y alignment with respect to the beam. Below, we describe a method developed to dynamically determine the alignment of the RPS detectors during the RPS$_{\rm track}$ data collection period. As described in Sec.~\ref{sec:xical}, the resulting $\xi_{\bar p}^{\rm RPS}$ distribution is used to calibrate $\xi_{\bar{p}}^{\rm CAL}$. 

\begin{figure}
\begin{center}
\includegraphics[width=0.5\textwidth]{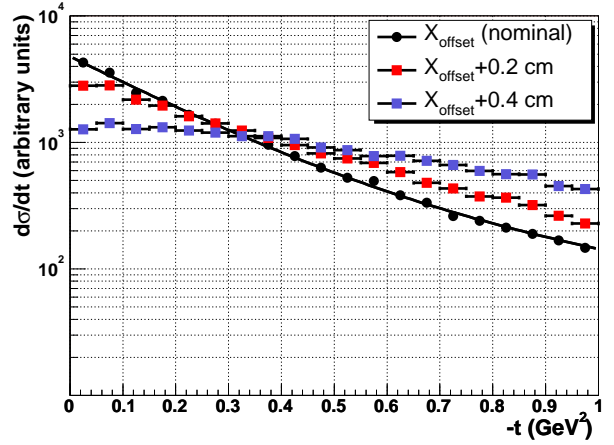}
\caption{$t$ distributions for SD RPS$_{\rm track}$ events of reconstructed RPS tracks in the range $0\le t\le 1$~GeV$^2$ for the nominal X position and for positive X$_{\rm offset}$ shifts relative to the beamline.}
\label{fig:rp_tpos} 
\end{center}
\end{figure}

The dynamic alignment method is illustrated in Fig.~\ref{fig:rp_tpos}, where the curve represents a fit to the data (after alignment) with a form composed of two exponential terms, 

\begin{equation}
\label{eq:tslope_fit}
\frac{d\sigma}{dt}=N_{norm}\left(A_1  e^{b_1t} +  A_2 e^{b_2 t}\right),
\end{equation}

\noindent where $N_{norm}$ is an overall normalization factor. 

\begin{figure}
\begin{center}
\includegraphics[width=0.5\textwidth]{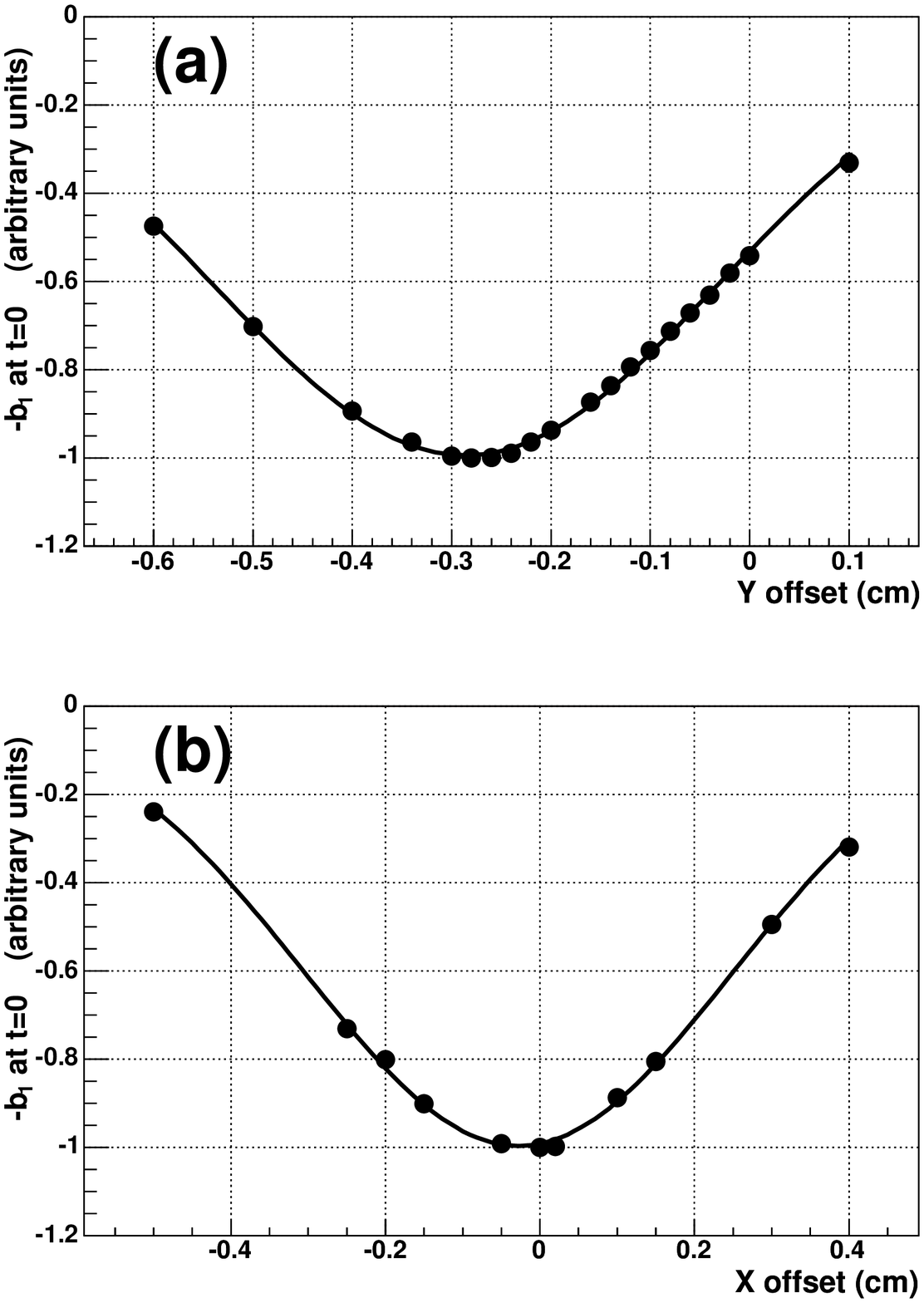}
\caption{\label{fig:offset} 
The value of $-b_1$ of the $t$-distribution parametrization of Eq.~(\ref{eq:tslope_fit})  normalized to its minimum value vs (a) Y$_{\rm offset}$ and (b) X$_{\rm offset}$.}
\end{center}
\end{figure}

Alignment is achieved  by seeking a maximum of the $d\sigma/dt$ distribution at $t=0$.  
The implementation of the alignment method consist of introducing software offsets X$_{\rm offset}$ and  Y$_{\rm offset}$ in the X and Y coordinates of the RPS detectors with respect to the physical beamline, and iteratively adjusting them until a maximum for $d\sigma/dt$  at $t=0$ (or of the dominant slope $b_1$) is found at the (X,Y) position where the RPS fiber tracker is correctly aligned.  
Results for such a fit are shown in Fig.~\ref{fig:offset}.
The accuracy in $\Delta \rm X$ and $\Delta \rm Y$ of the RPS alignment calibration obtained using the inclusive data sample, estimated from Gaussian fits to the distributions in Fig.~\ref{fig:offset} around their respective minimum values, is $\pm 60~\mu$m. This is limited only by the size of the data sample and the variations of the beam position during data taking. The contribution of the latter is automatically folded within the overall uncertainty as determined by this method.  

\subsection{Trigger efficiency of Roman pot spectrometer\label{RPS_trg_eff}}
Because of radiation damage, the position of the minimum ionizing particle (MIP) peak of the three RPT counters  shifts toward smaller ADC values as the integrated 
luminosity, $L$, of the data sample increases. The same behavior is observed in all three trigger counters.
As a direct consequence, the efficiency of the RPT triple coincidence decreases as a function of $L$.
The RPT efficiency, $\epsilon_{\rm RPT}$, is measured using a sample of minimum-bias (MB) data from 175.6 pb$^{-1}$ of integrated luminosity collected with a CLC$_p\cdot$CLC$_{\bar p}$ coincidence trigger.
The ADC distribution of each RPT counter was determined for various periods of data-taking by triggering with the other two RPT counters.
Results are shown in Table~\ref{tab:subsamples} for nine data sets of approximately equal integrated luminosity obtained by subdividing the MB data sample.   
We evaluate $\epsilon_{RPT}$ from the number of events with at least 1000 ADC counts in each of the three RPT counters (the trigger requirement)
divided by the number of events with at least 500 ADC counts in each counter, the lowest ADC value found among MIP peaks in all three counters and in all data sets: 

\begin{equation}
\epsilon_{RPT}=N_{1\cdot2\cdot3}({\rm ADC}> 1000) / N_{1\cdot2\cdot3} ({\rm ADC}> 500).
\label{eq:epsilonRPT}
\end{equation}

An uncertainty of 10\% is assigned to  $\epsilon_{RPT}$ to account for variations due to the choice of the lowest ADC value of the MIP peak as determined from an analysis of the all ADC distributions. The degradation of the RPT counters is taken into account by dividing the number of observed SD events by the RPT efficiency corresponding to the data-taking period to correct for RPT inefficiencies.

\begin{table}
\begin{center}
\caption{\label{tab:subsamples} Data sets of MB events collected with a CLC$_{\bar p}\cdot$CLC$_p$ coincidence requirement at various periods during the data-taking run, corresponding integrated luminosities, $L$, and 
triple-coincidence RPT counter efficiencies, $\epsilon_{\rm RPT}$.}
\vspace*{0.5em}
\begin{ruledtabular}
\begin{tabular}{ccc}
Data set & $L$ (pb$^{-1})$ & $\epsilon_{\rm RPT}$ \\ \hline
set~0 & 12.9 & $0.78\pm 0.08$ \\
set~1 & 24.0 & $0.75\pm 0.08$ \\
set~2 & 20.3 & $0.69\pm 0.07$ \\
set~3 &  6.4 & $0.57\pm 0.06$ \\
set~4 & 29.2 & $0.51\pm 0.05$ \\
set~5 & 16.3 & $0.46\pm 0.05$ \\
set~6 & 18.9 & $0.48\pm 0.05$ \\	
set~7 & 25.5 & $0.43\pm 0.04$ \\	
set~8 & 22.1 & $0.40\pm 0.04$ \\
\end{tabular}
\end{ruledtabular}
\end{center}
\end{table}

\subsection{Antiproton momentum loss measurement\label{sec:xisec}}

\begin{figure}[h]
\begin{center}
\includegraphics[width=0.5\textwidth]{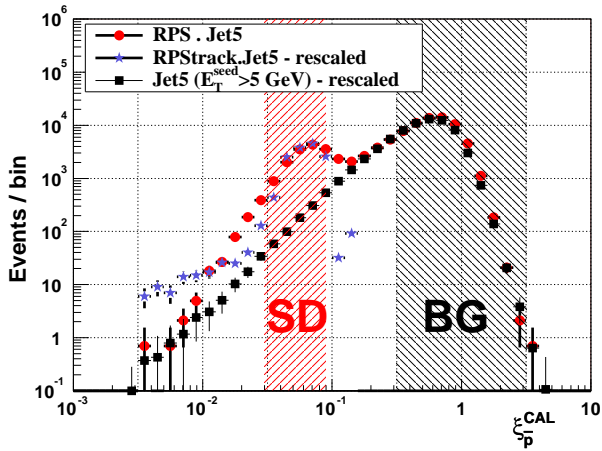}
\caption{\label{fig:Xi_withRPtrack}
The calorimeter-measured antiproton-momentum-loss fraction, $\xi_{\bar p}^{\rm{CAL}}$, for three data samples: (a) RPS$\cdot$Jet5 (filled circles), (b) RPS$_{\rm track}\cdot$Jet5-rescaled (filled stars), and (c) Jet5-rescaled (filled squares). The regions labeled SD and BG are respectively dominated by SD with no pileup and background events consisting of a ND dijet event and a SD event that provides the RPS trigger. The RPS$_{\rm track}\cdot$Jet5 and Jet5 distributions are normalized to the RPS$\cdot$Jet5 distribution in the SD and BG regions, respectively.}
\end{center}
\end{figure}

The momentum loss of the $\bar p$, $\xi_{\bar p}^{\rm CAL}$, is calculated using the pseudorapidity $\eta^i$  and transverse energy $E^i_T$ of all the towers of the CCAL, PCAL, and  MP calorimeters,

\begin{equation}
\label{eq:xi_formula}
\xi_{\bar p}^{\rm CAL}=\frac{1}{\sqrt{s}}\sum_{i=1}^{N_{\rm tower}}E_T^ie^{-\eta^i},
\end{equation}

\noindent where the sum is carried out over towers with $E_T>100$~MeV for CCAL and PCAL and $E_T>20$~MeV for MP.  Calibration issues of $\xi_{\bar pp}^{\rm CAL}$ are addressed in Sec.~\ref{sec:xical}.
The resulting $\xi_{\bar p}^{\rm CAL}$ distributions are shown in 
Fig.~\ref{fig:Xi_withRPtrack}.
The filled circles in this figure represent the $\xi_{\bar p}^{\rm CAL}$ distribution of the entire RPS$\cdot$Jet5 data sample
after applying all the selection 
requirements listed in Table~\ref{tab:SD_cuts}.
This distribution shows two peaks: one in the region of $\xi_{\bar p}^{\rm CAL}\lesssim 10^{-1}$ labeled SD, and a broader one at $\xi_{\bar p}^{\rm CAL}\gtrsim 10^{-1}$ labeled BG (background). 
The data points represented by filled squares are from the inclusive Jet5 data sample normalized (rescaled) to the number of the RPS$\cdot$Jet5 BG events in the region $ 3\times 10^{-1}< \xi^{\rm CAL}_{\bar p} <2$, and those represented by filled stars are from RPS$_{\rm track}\cdot$Jet5 data sample events with a reconstructed RPS track normalized (rescaled) to the number of RPS$\cdot$Jet5 events in the SD region of $3\times 10^{-2} < \xi^{\rm CAL}_{\bar p} < 9\times 10^{-2}$.

The RPS$\cdot$Jet5 data distribution is interpreted as follows:

\paragraph{Single diffractive region.} The events in the SD region are mainly due to SD dijet production.  The declining trend of the distribution below $\xi_{\bar p}^{\rm CAL}\sim 5\times 10^{-2}$ occurs in the region where the RPS acceptance is decreasing.
\paragraph{Background region.} The events in the BG region, which form the broad peak centered at $\xi_{\bar p}^{\rm CAL}\sim 0.5$, are mainly due to overlaps of a ND dijet event and a soft diffractive interaction that triggered the RPS but yielded no reconstructed vertex. A negligibly small fraction of the events in this region are due to SD dijet events superimposed with at least one MB event. Both types of overlap events should yield a value of $\xi_{\bar p}^{\rm CAL}\approx 1$, but the expected nearly $\delta$-function distribution is smeared by the resolution of the energy measurement in the calorimeters and shifted toward lower $\xi_{\bar p}^{\rm CAL}$-values by particles escaping detection 
either in the areas of the calorimeter interfaces or due to the imposed energy thresholds. 

As the rate of overlap events increases with instantaneous luminosity $\cal{L}$, 
the ratio of BG/SD events in the $\xi_{\bar p}^{\rm CAL}$ distribution of the RPS$\cdot$Jet5 event sample
 is expected to decrease with decreasing $\cal{L}$. We verified this effect in the RPS$\cdot$Jet5 data when binned in intervals of different $\cal{L}$.

\paragraph{Diffractive events.} The excess of the RPS$\cdot$Jet5 events over the Jet5 distribution in the SD region is mainly due to diffractive production with no pileup. The fraction of ND events in the region of $0.03<\xi_{\bar p}^{\rm CAL}<0.09$ is $\approx 12$\%. As the RPS acceptance depends on both $\xi_{\bar p}$ and $t$, and $t$ is not measured for each event, the background from ND events in the SD region is accounted for in a simple MC simulation designed to calculate the acceptance on an event-by-event basis.  Inputs to this simulation are the $\xi$ and $t$ distributions measured from data in which RPS tracking is available. Each event generated within a given bin of $\xi^{\rm RPS}$ is weighted by a factor equal to the ratio of the total number of events over the number of SD signal events in the corresponding $\xi_{\bar p}^{\rm CAL}$ bin of the RPS$\cdot$Jet5 data plotted in Fig.~\ref{fig:Xi_withRPtrack}.    



\subsection{Multiple interactions\label{sec:mult_interactions}}
Effects of multiple interactions in the same bunch crossing are handled by using the fraction of one-interaction events in the data sample, namely the fraction of events originating from bunch crossings that have just one $\bar pp$ interaction, to normalize the SD/ND event ratio in a multiple interaction environment. 
This fraction is estimated from a {\sc pythia}-Monte-Carlo-generated sample of events containing the appropriate run-dependent fraction of multiple interactions for each run. 
The instantaneous luminosities of the various subsamples comprising this dataset vary within the range $\sim 2<{\cal{L}}<4\times 10^{31}{\rm cm}^{-2}{\rm s}^{-1}$.
The ratio of the number of one-interaction events, $N^{\rm 1\_int}$, to all events, $N^{\rm all}$, decreases with increasing $\cal{L}$ following a distribution well 
described by an exponential expression

\begin{equation}
\label{eq:InstLum}
f^{\rm 1\_int}\equiv\frac{N^{\rm 1\_int}}{N^{\rm all}}\propto\exp[-c\cdot {\cal{L}}],
\end{equation}

\noindent where $c=(0.34\pm0.06)\times 10^{-31}\;{\rm cm}^2{\rm s}$.

An offline analysis vertex requirement (cut) accepting events with only one reconstructed vertex would eliminate a large fraction of ND overlap events, but might also reject diffractive events due to vertex reconstruction inefficiencies, which depend on the event activity and therefore on both $\xi$ and the number of overlapping events. To avoid biasing the $\xi_{\bar p}^{\rm CAL}$ distribution,  we apply no vertex cut and correct for the ND overlap event contamination when we evaluate cross sections. 

\subsection{Calibration of $\xi^{\rm CAL}_{\bar p}$\label{sec:xical}}
\begin{figure}[h]
\begin{center}
\includegraphics[width=0.5\textwidth]{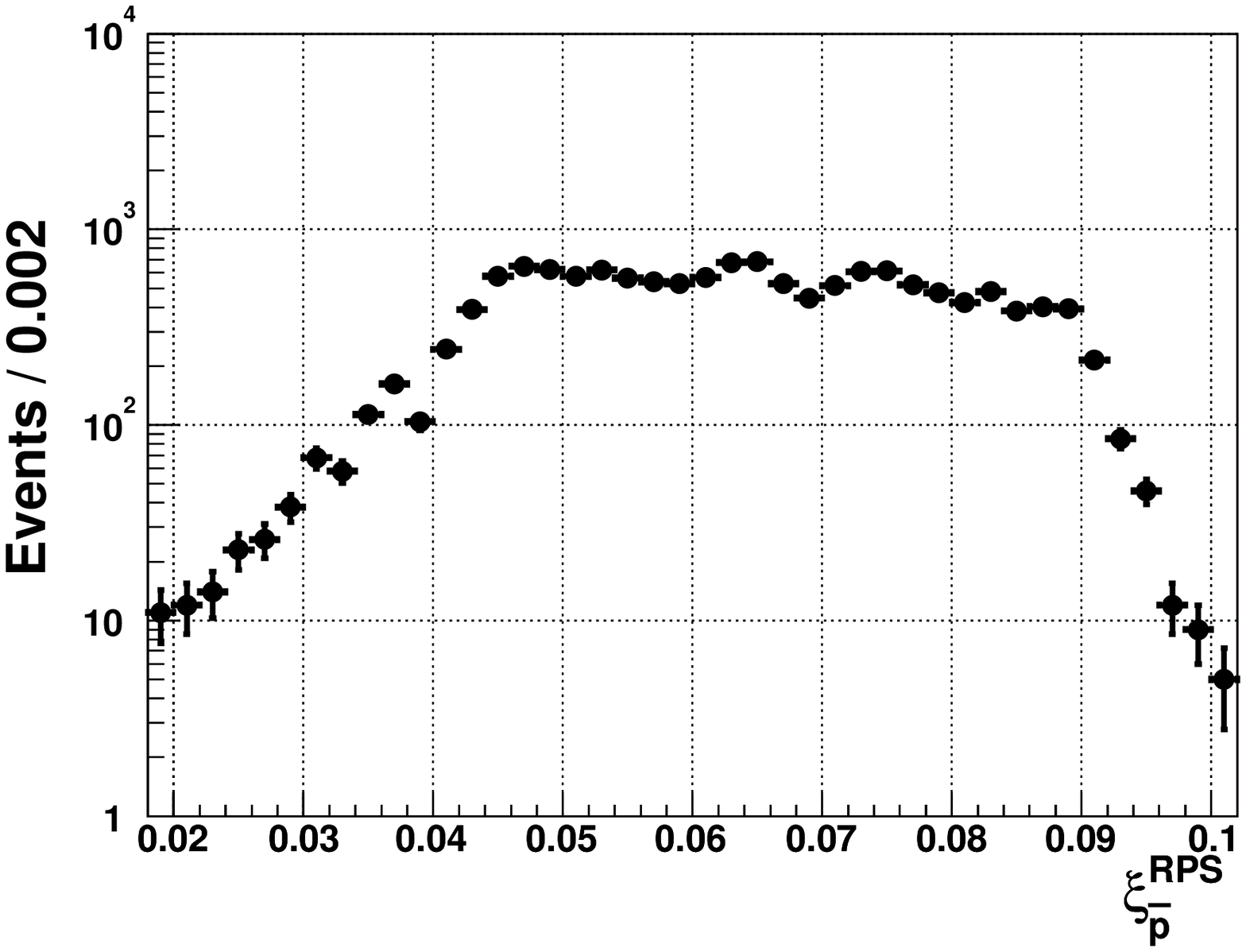}
\caption{\label{fig:xi_rp} 
The $\xi^{\rm RPS}_{\overline{p}}$ distribution of the RPS$_{\rm track}$ events.}
\end{center}
\end{figure}

The measurement of $\xi_{\bar p}^{\rm CAL}$ using Eq.~(\ref{eq:xi_formula})
is calibrated by comparing with the RPS  measurement of $\xi_{\bar p}^{\rm RPS}$. The comparison is made using the RPS$_{\rm track}$ subsample of events for which tracking information is available.   
The $\xi_{\bar p}^{\rm RPS}$ distribution for these events is shown in Fig.~\ref{fig:xi_rp}. 

\begin{figure}
\begin{center}
\includegraphics[width=0.5\textwidth]{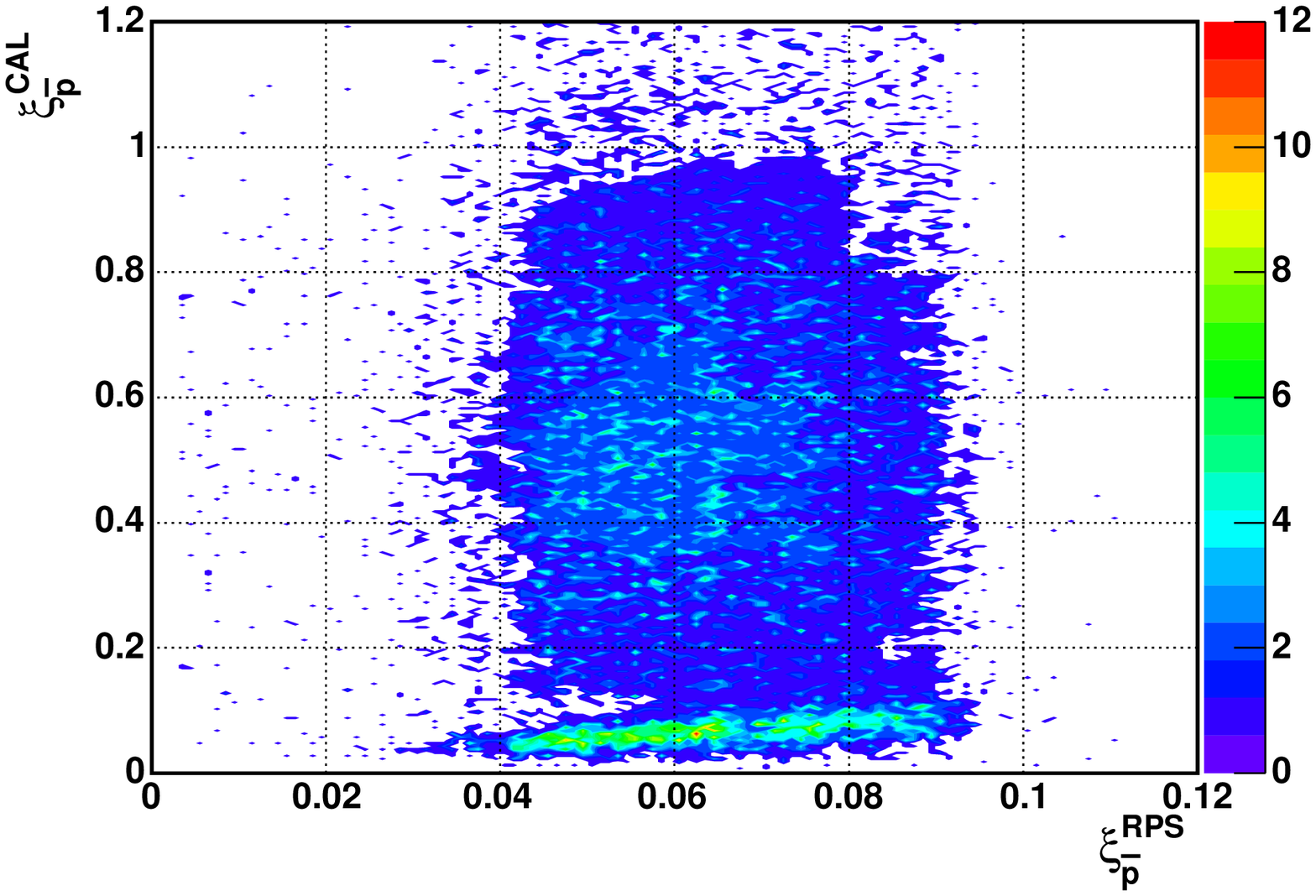}
\caption{\label{fig:xi_rp_vs_mp} 
Scatter plot of $\xi_{\overline{p}}^{\rm CAL}$ vs $\xi_{\overline{p}}^{\rm RPS}$ for all RPS$_{\rm track}$ data.}
\end{center}
\end{figure}

\begin{figure}
\begin{center}
\includegraphics[width=0.5\textwidth]{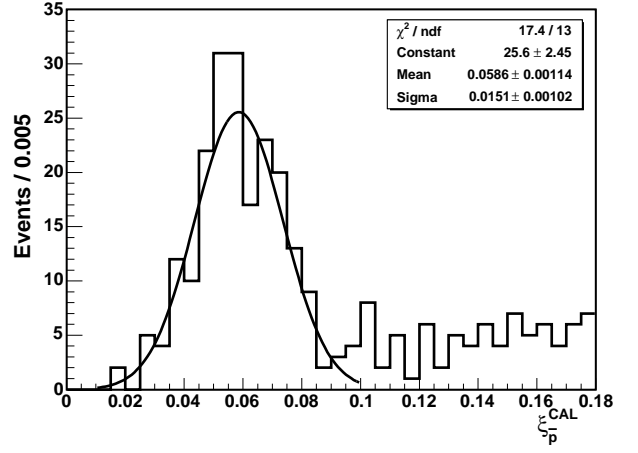}
\caption{\label{fig:xi_rp_vs_mp_sample} 
The $\xi^{\rm CAL}_{\overline{p}}$ distribution of RPS$_{\rm track}$ events in the range $0.055<\xi^{\rm RPS}_{\overline{p}}<0.060$. The curve is a Gaussian fit in the region of  $\xi^{\rm CAL}_{\overline{p}}<0.1$.}
\end{center}
\end{figure}

\begin{figure}[htp]
\begin{center}
\vspace*{-0.8cm}
\includegraphics[width=0.49\textwidth]{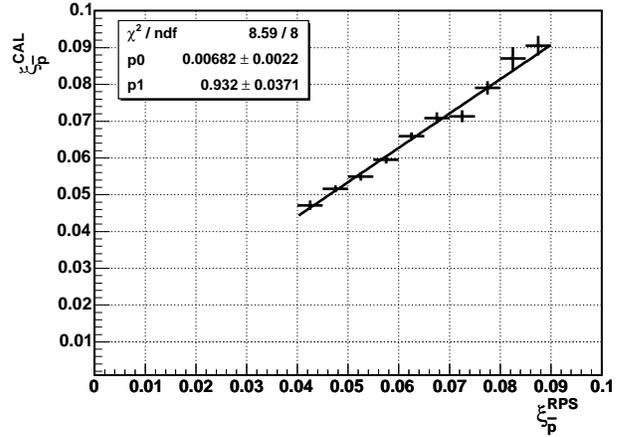}
\caption{Median values of $\xi^{\rm CAL}_{\overline{p}}$ obtained from fits to RPS$_{\rm track}$ data in different $\xi^{\rm CAL}_{\overline{p}}$ bins of width $\Delta\xi^{\rm CAL}_{\overline{p}}=0.05$ (horizontal bars) fitted linearly in the region $0.045<\xi^{\rm CAL}_{\overline{p}}<0.09$. }
\label{fig:xi_rp_vs_mp_linfit} 
\end{center}
\end{figure}

\begin{figure}
\begin{center}
\includegraphics[width=0.5\textwidth]{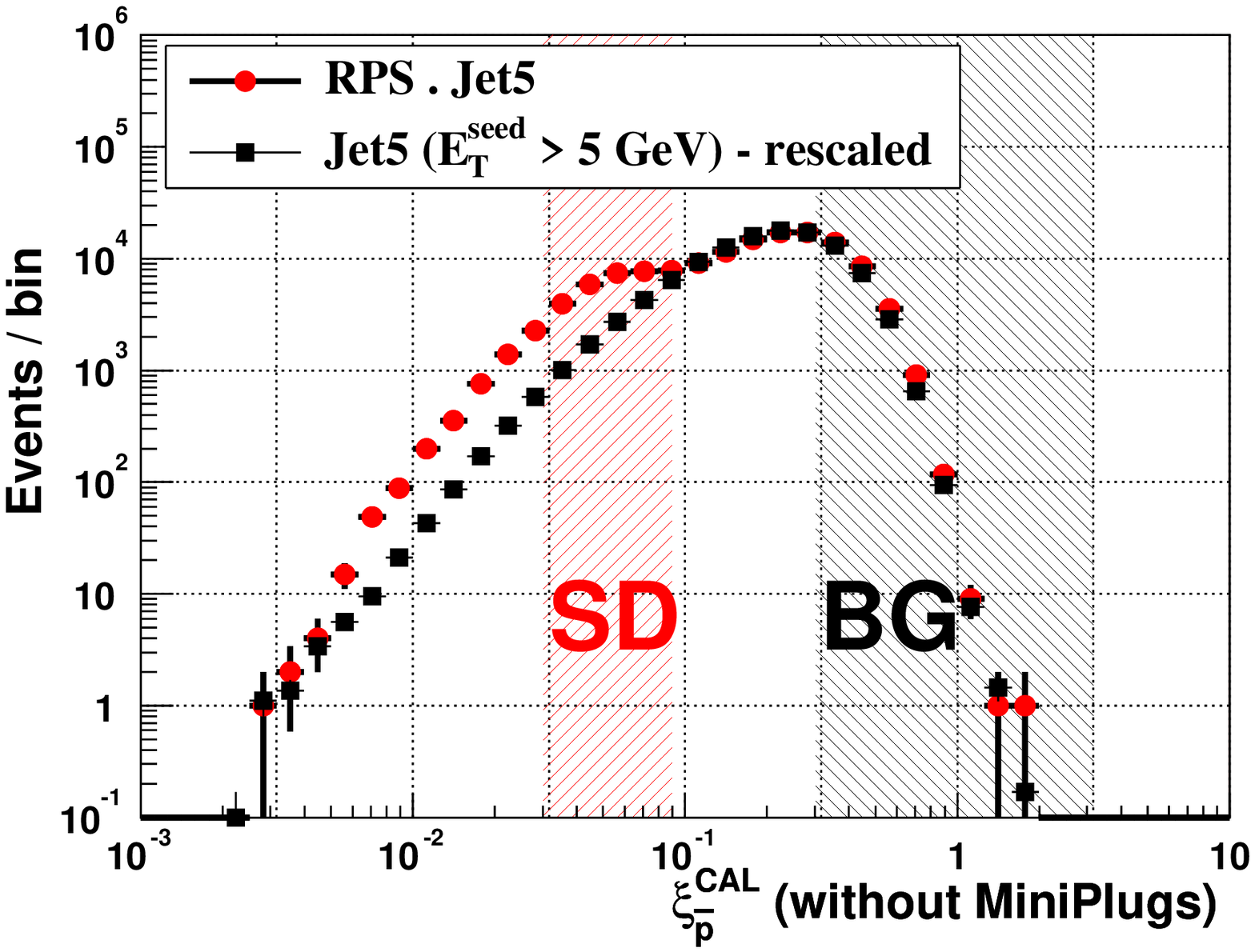}
\caption{\label{fig:MPCal2} The $\xi_{\overline{p}}^{\rm CAL}$ distribution for RPS$\cdot$Jet5 and Jet5 data evaluated without the MiniPlug calorimeter contributions. The diffractive peak is less discernible in the RPS$\cdot$Jet5 distribution, and the Jet5 background is larger than in Fig.~\ref{fig:Xi_withRPtrack} where the MiniPlugs are included in the evaluation of $\xi_{\overline{p}}^{\rm CAL}$.}
\end{center}
\end{figure}

Figure~\ref{fig:xi_rp_vs_mp} shows a two-dimensional scatter plot of $\xi_{\bar p}^{\rm CAL}$ vs $\xi_{\bar p}^{\rm RPS}$ for the events with a reconstructed RPS track. An approximately linear relationship between $\xi_{\bar p}^{\rm CAL}$ 
and $\xi_{\bar p}^{\rm RPS}$ is observed along the peak of the distribution of entries in the region of 
$\xi_{\bar p}^{\rm CAL}\lesssim 0.1$ and $0.03\lesssim\xi_{\bar p}^{\rm RPS}\lesssim 0.09$, 
where the diffractive events are expected.
As already discussed, the entries with $\xi_{\bar p}^{\rm CAL}> 0.1$ are mainly due to a ND dijet event with a superimposed soft SD overlap event. 
Without the measurement of $\xi_{\bar p}^{\rm CAL}$ the overlap events would contribute a large and irreducible background to the SD events.

For a quantitative calibration of $\xi_{\bar p}^{\rm CAL}$
the data are divided into bins of width $\Delta\xi_{\bar p}^{\rm RPS}=0.005$ and the $\xi_{\bar p}^{\rm CAL}$ 
values in each bin are fit with a Gaussian distribution.
Figure~\ref{fig:xi_rp_vs_mp_sample} shows such a fit for the bin of $0.055<\xi_{\bar p}^{\rm CAL}<0.060$. 
The ratio of the half-width to the average value of the fit curve is found to have the constant value of $\delta\xi_{\bar p}^{\rm CAL}/\xi_{\bar p}^{\rm CAL}\approx 0.3$ over the entire $\xi_{\bar p}^{\rm CAL}$ range of this data-sample.

The results of $\xi_{\bar p}^{\rm CAL}$ vs $\xi_{\bar p}^{\rm RPS}$ 
for all the $\Delta\xi_{\bar p}^{\rm CAL}$ bins 
 within the plateau region of $0.045<\xi_{\bar p}^{\rm RPS}<0.09$  of the $\xi_{\bar p}^{\rm CAL}$ distribution of Fig.~\ref{fig:xi_rp} are plotted in Fig.~\ref{fig:xi_rp_vs_mp_linfit}. A linear relationship is observed of the form $\xi_{\bar p}^{\rm CAL}=p^0+p^1\cdot \xi_{\bar p}^{\rm RPS}$ with $p^0=0.007\pm0.002$ and 
$p^1= 0.97\pm 0.04$. The small size of the deviations of $p^0$ and $p^1$ from the ideal values of $p^0=0$ and $p^1=1$ is testimony to the success of the dynamic alignment of the RPS, which affects $\xi_{\bar p}^{\rm RPS}$, and the calorimeter energy calibrations and threshold settings, which influence $\xi_{\bar p}^{\rm CAL}$. These deviations do not affect the referral of the measurement of $\xi_{\bar pp}^{\rm CAL}$ to that of  $\xi_{\bar pp}^{\rm RPS}$ to obtain an RPS-based value of $\xi_{\bar p}$ in the data for which no RPS tracking is available.    

\subsection{MiniPlug contribution to $\xi_{\bar p}^{\rm CAL}$}

The measured value of $\xi_{\bar p}^{\rm CAL}$ receives contributions from all the CDF calorimeter subsystems.
The largest contribution at large $\xi_{\bar p}$ comes from 
PCAL$_{\bar p}$ and MP$_{\bar p}$, the PCAL and MiniPlug calorimeters on the $\bar{p}$ side. The contribution of the MP$_{\bar p}$ is particularly important for reducing the background due to ND dijet events overlapped by a soft SD recoil $\bar p$ detected in the RPS.
Figure~\ref{fig:MPCal2} shows the $\xi_{\bar p}^{\rm CAL}$ distribution for the RPS$\cdot$Jet5 and Jet5 data without the contribution from the MP calorimeters.  In comparison with Fig.~\ref{fig:Xi_withRPtrack}, the BG peak is shifted towards the SD region and contributes a considerably larger ND background to the SD events.

The MP calibration was performed by comparing $E_T$ distributions between  data and Monte Carlo generated events. The slopes of the $dN/dE_T$ distributions of the data were first equalized among the MP towers within a given polar angular range ($\eta$-ring), and then normalized to the corresponding slopes obtained from Monte Carlo generated events in the same $\eta$-ring.
The resolution in the measurement of $E_T$ was estimated to be $\delta E_T=\pm 30\%$ using data obtained with a MP prototype calorimeter exposed to high energy positron and pion beams~\cite{MP_prototype_test}  . 
Varying the MP energy by $\pm$~30\% yields a shift of $\Delta\log_{10}\xi_{\bar p}^{\rm CAL}=\pm 0.1$, which is comparable to the bin width  used in the analysis (see the data plotted in Fig.~\ref{fig:Xi_withRPtrack}). 

\subsection{Beam-halo background\label{beamhalo}}

The background in the $\xi_{\bar p}^{\rm CAL}$ distribution due to beam-halo (particles at relatively large distances off the beam axis) was studied using a trigger provided by the Tevatron clock at the nominal time the $\bar p$ and $p$ beam bunches cross the $z=0$ position at the center of CDF~II regardless of whether or not there is a $\bar pp$ interaction. This trigger, which is referred to as ``zero-crossing,'' leads to various data samples depending on additional conditions that may be imposed. 

Three zero-crossing event samples were analyzed: 

(a) zero-crossing inclusive;

(b) zero-crossing events with no reconstructed vertex; 

(c) zero-crossing events with a RPS trigger. 

\begin{figure}
\begin{center}
\includegraphics[width=0.5\textwidth]{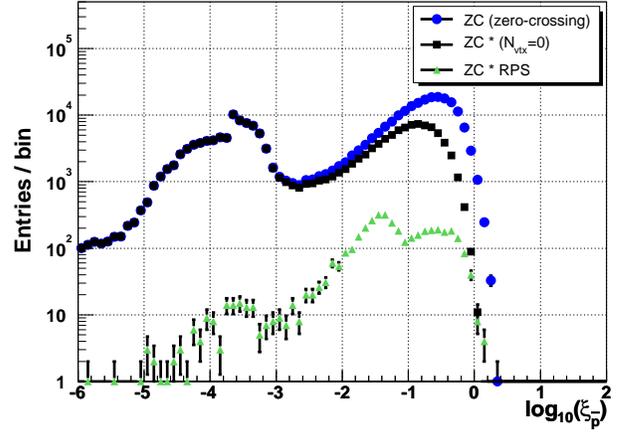}
\caption{The $\xi_{\overline{p}^{\rm CAL}}$ distribution in zero-crossing (ZC) events (circles), ZC with no reconstructed vertices (squares), and
ZC  with a RPS trigger (triangles).}
\label{fig:xi_RPbkg} 
\end{center}
\end{figure}

Figure~\ref{fig:xi_RPbkg} shows the $\xi_{\bar p}^{\rm CAL}$ distribution for these samples. Three regions of $\xi_{\bar p}^{\rm CAL}$ are of interest:
\begin{enumerate}[(i)]
\addtolength{\itemsep}{-0.5em}
\item $\xi_{\bar p}^{\rm CAL}>0.1$, where ND events dominate;
\item $10^{-3}<\xi_{\bar p}^{\rm CAL}<0.1$, where SD events are expected;
\item  $\xi_{\bar p}^{\rm CAL}<10^{-3}$, where ``empty'' events with one to a few  CCAL or PCAL towers above threshold due to beam-halo particles and/or due to calorimeter noise may contribute to $\xi_{\bar p}^{\rm CAL}$.
\end{enumerate}

The contribution of a single CCAL or PCAL tower at $\eta=0$ with $E_T=0.2$~GeV (the threshold used) is estimated from Eq.~\ref{eq:xi_formula} 
to be $\xi_{\bar p}^{\rm CAL}=1\times 10^{-4}$, which is well below the $\xi$-range of the RPS acceptance. At $\xi_{\bar p}^{\rm CAL}\sim 3\times 10^{-4}$,  we observe a background peak corresponding to an average of $\sim 3$ CCAL towers at an event rate of $\sim 5\%$ of the diffractive signal concentrated at $\xi_{\bar p}^{\rm CAL}\sim 5\times 10^{-2}$. We estimate that an upward fluctuation by a factor of 100 would be required for this background to compete with the overlap background already present within the diffractive region of $3\times 10^{-2}<\xi_{\bar p}^{\rm CAL}<9\times 10^{-2}$. Because of the negligible probability of such a fluctuation, no correction is applied to the data for beam-halo background. 

\section{Results}\label{results}
\begin{figure}[htp]
\begin{center}
\includegraphics[width=0.5\textwidth]{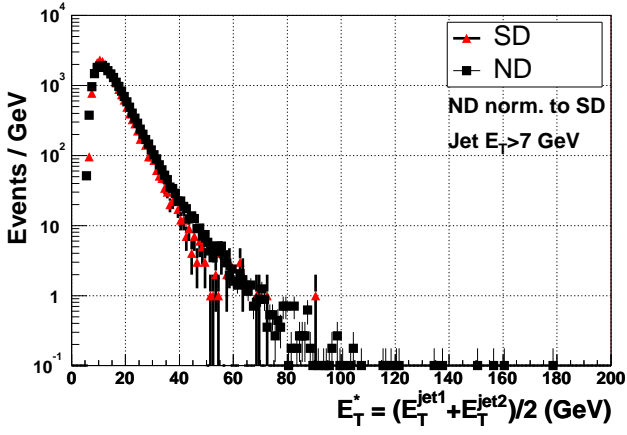}
\caption{\label{fig:JetEtAve} 
The distribution of the mean dijet transverse energy $E_T^*=(E_T^{\rm jet1}+E_T^{\rm jet2})/2$  
for SD and ND events. The ND distribution is normalized to the total number of SD events.}
\end{center}
\end{figure}

\begin{figure}[htp]
\begin{center}
\includegraphics[width=0.5\textwidth]{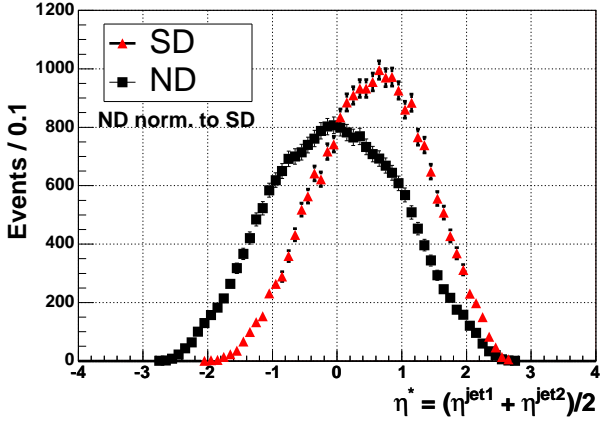}
\caption{\label{fig:JetDetaAve} 
The average $\eta$ distribution, $\eta^*=(\eta^{\rm jet 1}+\eta^{\rm jet 2})/2$, of the two highest $E_T$ jets for SD and ND events. The ND distribution is normalized to the total number of SD events.}
\end{center}
\end{figure}

\begin{figure}[htp]
\begin{center}
\includegraphics[width=0.48\textwidth]{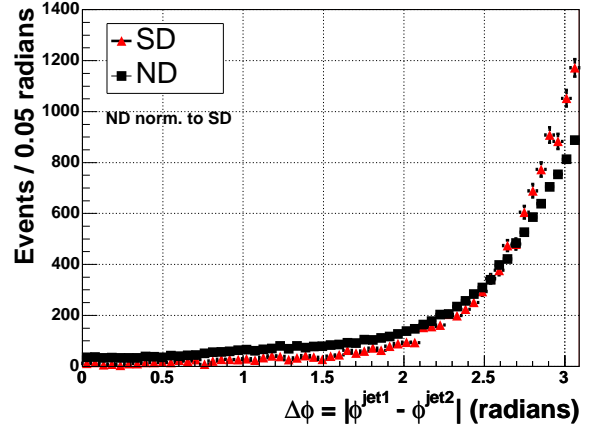}
\caption{\label{fig:rpj5_dphi} 
The distribution of the azimuthal angle difference $\Delta\phi$ between the two highest $E_T$ jets for SD and ND events. The ND distribution is normalized to the total number of SD events.}
\end{center}
\end{figure}

\begin{figure}[htp]
\begin{center}
\vspace*{-0.5cm}
\includegraphics[width=0.5\textwidth]{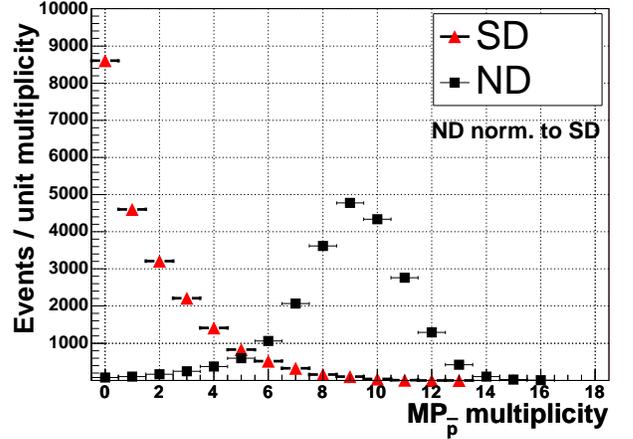}
\caption{\label{fig:mpw_mult} 
The multiplicity distribution in the MP$_{\bar p}$ calorimeter for SD and ND events. The ND distribution is normalized to the total number of SD events.}
\end{center}
\end{figure}

Our results are presented in  three subsections~\ref{kinematic}-\ref{tdist}. In \ref{kinematic}, we discuss certain kinematic distributions that establish the diffractive  nature of the events
 in our data samples; in \ref{ratio}, we present ratios of SD to ND production rates and extract the diffractive structure function; and in~\ref{tdist}, we report results on $t$ distributions.  
     
\subsection{Kinematic distributions}\label{kinematic}
The presence of a rapidity gap in diffractive events leads to characteristic kinematic distributions. Here, we compare the SD and ND distributions of mean transverse jet energy, mean jet pseudorapidity, jet azimuthal angle difference, and MiniPlug multiplicity.

\setcounter{paragraph}{0}
\paragraph{Transverse jet energy.} The mean dijet transverse energy, $E_T^*=(E_T^{jet1}+E_T^{jet2})/2$, is presented in Fig.~\ref{fig:JetEtAve} for SD and ND events. 
The total number of ND events is normalized to that of the SD events and the associated statistical uncertainties are rescaled. The SD and ND distributions are very similar. The slightly narrower width of the SD distribution is attributed to the lower effective collision energy in SD ($\Pomeron$-$p$ collision), and the larger fraction of jets at $\eta>0$ in SD compared to ND  events, which tend to have a lower average $E_T$~\cite{ken_thesis}. 

\paragraph{Jet pseudorapidity.} The mean dijet pseudorapidity distribution of the two leading jets, $\eta^* = (\eta^{\rm jet1} + \eta^{\rm jet2}) /2$, 
is shown for SD and ND events in Fig.~\ref{fig:JetDetaAve}. The ND events are centered around $\eta^* =0$, while the SD distribution is shifted towards positive values (proton-beam direction) due to the boost of the center of mass system of the $\Pomeron$-$p$ collision. 

\paragraph{Azimuthal angle correlations.} The distributions of the azimuthal angle difference between the two leading jets, $\Delta{\rm \phi} =|{\rm \phi^{jet1}} - \rm {\phi^{jet2}}|$, are shown in  Fig.~\ref{fig:rpj5_dphi} for SD and ND events. The SD dijets are more back-to-back than the ND ones, as would be expected from the reduced available subenergy of the diffractive system.  

\paragraph{MiniPlug multiplicity.} The MP$_{\bar p}$ multiplicity distribution for SD and ND events is shown in Fig.~\ref{fig:mpw_mult}. Multiplicities are evaluated by counting the number of peaks above calorimeter tower noise levels using ``seed'' towers with a minimum transverse energy of 20~MeV~\cite{michele_rio}. Such peaks originate from three sources: (a) charged hadrons traversing the MP without interacting, (b) charged and neutral hadrons interacting in the material of the MP plates, and (c) electromagnetic energy deposited by $e^\pm/\gamma$ particles.      
The SD events have smaller multiplicities in comparison with the ND events, as expected from the reduced center of mass energy in $\Pomeron$-$p$  relative to $\bar p$-$p$ collisions.

\subsection{Ratio of single diffractive to nondiffractive production rates and the diffractive structure function\label{ratio}}

The ratio ${\cal{R}}\equiv {\rm R}_{\rm SD/ND}(x, Q^2, \xi, t)$ of the SD to ND dijet production rates, which in leading order QCD is proportional to the ratio of the corresponding structure functions~(see Sec.~\ref{method}), is  measured as a function of $x_{Bj}^{\bar p}$, the Bjorken-$x$ of the struck parton of the antiproton, and $Q^2\approx \langle E_T^*\rangle^2$. 
For each event, $x_{Bj}^{\bar p}$ is evaluated from the $E_T$ and $\eta$ values of the jets using the formula

\begin{equation}
x_{Bj}^{\bar p}=\frac{1}{\sqrt{s}}\sum_{i=1}^{\rm 3~jets}E_T^ie^{-\eta^i},
\label{eq:xbj}
\end{equation}

\noindent
where the sum is carried out  over the two leading jets plus a third jet of $E_T>5$~GeV, if present. Theoretically, the sum should be over all jets in the final state, but the fraction of events with more than three jets of $E_T>5$~GeV is relatively small and including them in the evaluation of $x_{Bj}^{\bar p}$ does not significantly affect the obtained results~\cite{ken_thesis}. 

Jet energies are measured using an algorithm based on measuring the ``visible'' energy deposited in the detector within a cone of radius $R_{\rm cone}=\sqrt{\delta\eta^2+\delta\phi^2}=0.7$ and applying appropriate corrections~\cite{jetcorrections}. Both the SD and ND jets contain within the jet cone an amount of underlying event (UE) energy from soft spectator partonic interactions, which we subtract from the measured jet energy. For diffractive events, where a large fraction of the event energy is carried away by the recoil antiproton, the amount of UE energy is expected to be smaller than in ND events.

The UE energy in SD events was measured using the sample of RPS inclusive triggered data. To suppress overlap backgrounds, we required the events to be in the region of
$0.01 < \xi_{\bar{p}}^{\rm CAL} < 0.14$, which is dominated by SD events from a single $\bar{p} p$ interaction. 
We then selected events with only one reconstructed vertex 
with $|z_{\rm vtx}|<60$~cm ($\sim 1
\,\sigma$) and measured the $\Sigma E_T$ of all central calorimeter towers within a randomly chosen cone of radius $R_{\rm cone}=0.7$ in the region $0.1 < |\eta| < 0.7$. 
The UE energy in MB (ND dominated) events was also measured using the random cone technique.
From these two  measurements we obtained an average UE $E_T$ of 0.90~GeV (1.56~GeV) for SD (ND) events.

\subsubsection{$x_{Bj}$ dependence}\label{xbj_dependence}

The ratio $\cal{R}$ of the number of SD dijet events per unit $\xi$ over the number of ND events of the Jet5 sample is corrected for the effect of multiple interactions and for the RPS detector acceptance. 

To account for multiple interactions contributing to the ND  sample, the ND $x_{Bj}$ distribution is weighted by the factor $f^{\rm 1\_int}=\exp[-c\cdot {\cal{L}}]$, where $\cal{L}$  is the instantaneous luminosity and $c=(0.34\pm0.06)\times 10^{-31}\;{\rm cm}^2{\rm s}$ (see Sec.~\ref{sec:mult_interactions}). This correction is not applied to the SD events, since contributions from additional interactions shift $\xi^{\rm CAL}_{\bar p}$ into the BG region. The effect of overlaps of a SD and a ND event with jets in both events, which would tend to shift the $x_{Bj}$ distribution towards higher values, is estimated to be on the 0.1\% level and is neglected.

The number of SD events is corrected for the RPS acceptance, which is estimated from a beam optics simulation to be $\epsilon_{\rm}=0.80\pm 0.04\;{\rm (syst)}$ for $0.03<\xi<0.09$ and $t >1$~GeV$^2$~\cite{excl2j}, and the obtained SD and ND data samples are normalized to their respective integrated luminosities.

\begin{figure}[h]
\begin{center}
\includegraphics[width=0.5\textwidth]{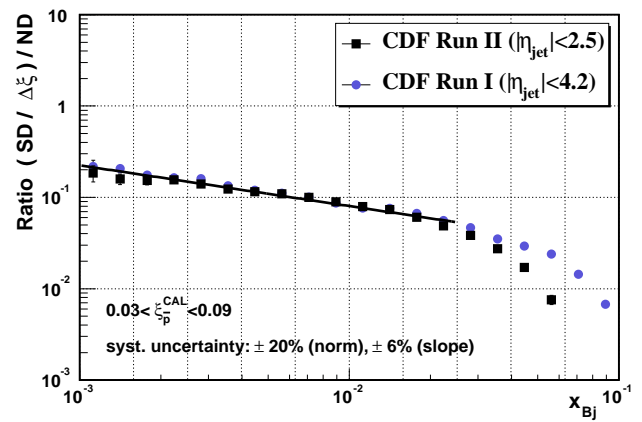}
\caption{\label{fig:diff_sf_sd} 
The ratio of single-diffractive (SD) to nondiffractive (ND) dijet-event rates as a function of $x_{\rm Bj}$. The quoted overall systematic uncertainty of $\pm 20$\% for the Run~II ratio is due to the uncertainties $\Delta{\cal{R}}_0$ and $\Delta r$ listed in Table~\ref{tab:systematics}. The overall systematic uncertainty in the Run~I ratio is $\pm 25$\%.}
\end{center}
\end{figure}

The results for the ratio $\cal{R}$ for events within $0.03<\xi<0.09$ are presented in Fig.~\ref{fig:diff_sf_sd}  as a function of $\log_{10} x_{\rm Bj}$, along with the results obtained in Run~I for $0.035<\xi<0.095$ and $|t|<1$~GeV$^2$~\cite{jjRPS}. The Run~I results have an estimated $\pm 25\%$ systematic uncertainty in the overall normalization~\cite{jjRPS}. The shapes of the  two distributions are in good agreement, except in the high $x_{Bj}$ region where the Run~I distribution is seen to extend to higher $x_{Bj}$ values. This behavior is expected from the difference in $\langle E_T^*\rangle$ and $\eta$ acceptance for jets in Run~I (Run~II), which is $\langle E_T^*\rangle\approx 7$~GeV (12~GeV) and $|\eta|<3.7$ ($|\eta|<2.7$). This difference results in a larger $x_{Bj}$ reach in Run~I by a factor of $(12/7)\times e^{3.7-2.7}\approx 2$, as estimated using Eq.~\ref{eq:xbj}.

A fit to all Run~II data in the range $0.03<\xi<0.09$ using the form ${\cal R} = {\cal R_{\rm 0}}\cdot x_{Bj}^{r}$ subject to the constraint $\beta=(x/\xi)<0.5$~\cite{jjRPS} yields $r=-0.44\pm0.04$ and ${\cal R_{\rm 0}}= (8.6\pm 0.8)\times10^{-3}$. This result is compatible with the Run~I result of $r=-0.45\pm 0.02$ and ${\cal R_{\rm 0}}= (6.1\pm 0.1)\times10^{-3}$ obtained by a constrained fit of the form ${\cal R} = {\cal R_{\rm 0}}\cdot (x_{Bj}/0.0065)^{r}$ designed to match the data at the average value of $x_{Bj}/0.0065$. This constraint accounts for the lower ${\cal R_{\rm 0}}$ value of Run~I.

{\bf The systematic uncertainties in ${\cal R_{\rm0}}$ and $r$} are listed in Table~\ref{tab:systematics}. The causes of uncertainty investigated include the underlying event, the energy scale, the calorimeter tower $E_T$ thresholds, overlaps in ND events, instantaneous luminosity $\cal{L}$, bunch-by-bunch variations in $\cal{L}$, RPS acceptance, and ``splash'' events. 
\paragraph{Underlying event.}
The underlying event energy is subtracted from the jet energy when the jet energy corrections are applied. The results presented are for a $\pm$30\% variation of the UE energy correction, which is sufficient to cover the uncertainty for jets depositing energy near the interfaces or the outer edges of a calorimeter.
\paragraph{Energy scale.}
The effect of the energy scale of the CCAL, PCAL and MP calorimeters on $\xi_{\bar p}^{\rm CAL}$ and thereby on ${\cal R_{\rm 0}}$ and $r$ was estimated by changing the CCAL and PCAL jet energies by $\pm 5$\% and the MP tower energies by $\pm 30$\%, based on studies of inclusive jets and comparisons of MP $E_T$ distributions with expectations based on simulations.
\paragraph{Tower $E_T$ threshold.}
Tower energy threshold effects would generally be expected to cancel out in measuring the ratio of rates.
However, due to the different UE event contributions in SD and ND events, 
the tower thresholds applied could affect the result.
The uncertainty in ${\cal R}$ for jets with mean transverse energy of $E_T^*>10$~GeV ($E_T^*>12$~GeV) 
due to tower threshold effects is estimated to be +1\% (+2\%). The effect on both ${\cal R_0}$ and on $r$ for our total SD event sample is $\pm 1\%$.
\paragraph{Overlaps in ND events.}
To account for event overlaps in ND events occurring at high instantaneous luminosities, each event is weighted by a factor of $w=\exp[-c\cdot\cal{L}]$, 
where $c=(0.34\pm0.06)\times 10^{-31}\;{\rm cm}^2{\rm s}$ (see Sec.~\ref{sec:mult_interactions}). 
The uncertainty in $c$ has a $\pm 8\%$ effect on the determination of ${\cal R_{\rm 0}}$.
\paragraph{Instantaneous luminosity, $\cal{L}$.}
We estimate an uncertainty of $\pm3\%$ in the determination of ${\cal R_{\rm 0}}$ due to an uncertainty of $\pm 6\%$ in the determination of $\cal{L}$ during the data collection period.
\paragraph{Bunch-by-bunch variation in $\cal{L}$.}
A bunch-by-bunch variation of $\sim$50\% in $\cal{L}$ over the collection period of our data sample contributes a $\pm 4\%$ systematic uncertainty in ${\cal R_{\rm 0}}$.
\paragraph{RPS acceptance.}
Using a MC simulation, the uncertainty in ${\cal R_{\rm 0}}$ as estimated to be $\pm 10\%$.
\paragraph{Splash events.}
By evaluating the differences among data subsamples, a systematic uncertainty of $\pm 6\%$ is estimated due to the cut applied to remove the splash events (see Table~\ref{tab:q2_summary}).

Added in quadrature, the above uncertainties yield $\Delta{\cal{R}}\mbox{ (syst)}=\pm18\%$ and $\Delta r\mbox{ (syst)}=\pm6\%$. 
\begin{table}[htb]
\begin{center}
\caption{\label{tab:systematics} Systematic uncertainties $\Delta{\cal R}_0$ and $\Delta r$ in ${\cal R}_0$ and $r$ of the ratio of SD to ND events parametrized as ${\cal R} = {\cal R}_0\cdot x_{Bj}^r$.  The uncertainties are  estimated either by varying the parameter associated with the source of error by an amount denoted by ``Variation,'' or in the cases marked as ``$N/A$'' (not applicable) by a direct calculation using data or simulated event distributions as explained in the text.}
\vspace*{0.5em}
\begin{tabular}{lllr}\hline\hline
Source of uncertainty				& Variation	& $\Delta{\cal R}_0$&	$\Delta r$\\ \hline
Underlying event				& $\pm$30\%	& $\pm$3\%	& $\mp$5\% \\ 
Central/plug cal. energy scale			& $\pm$5\%	& $\pm$8\%	& $\sim$3\% \\
MiniPlug energy scale 				& $\pm$30\%	& $\mp$6\%	& $\pm$1\% \\
Tower $E_T$ threshold				& $\pm$10\%	& $\pm$1\%	& $\pm$1\% \\ 
Overlaps in ND events					& $\pm$20\%	& $\pm$8\%	& $<\pm$1\%\\ 
Instantaneous luminosity			& $\pm$6\%	& $\pm$3\%	& N/A      \\ 
Bunch-by-bunch luminosity			& $\pm$50\%	& $\pm$4\%	& N/A      \\ 
RPS acceptance					& N/A       	& $\pm$10\%	& N/A	   \\
Splash events					& N/A       	& $\pm$6\%	& N/A      \\ \hline
Total uncertainty						&		& $\pm$18\%	& $\pm$6\% \\
\hline\hline 
\end{tabular}
\end{center}
\end{table}

\subsubsection{$Q^2$ dependence}
\begin{figure}
\begin{center}
\includegraphics[width=0.5\textwidth]{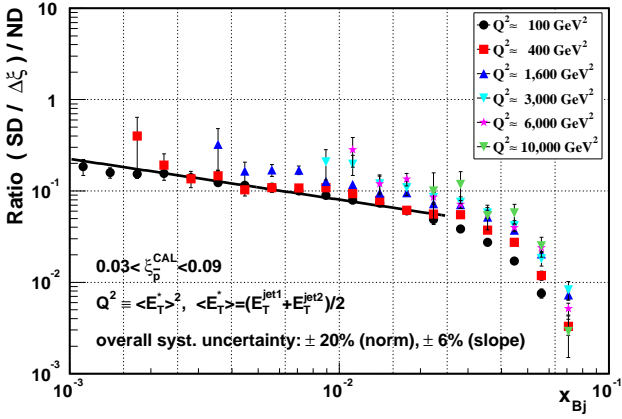}
\caption{\label{fig:sf_q2} 
The ratio of diffractive to nondiffractive dijet event rates as a function of $x_{\rm Bj}$ 
(momentum fraction of parton in the antiproton) for different values of $Q^2\approx\langle E_T^*\rangle^2$. The quoted overall systematic uncertainty of $\pm 20$\% is due to the uncertainties $\Delta{\cal{R}}_0$ and $\Delta r$ listed in Table~\ref{tab:systematics}.}
\end{center}
\end{figure}

As discussed in Sec.~\ref{data}, special data samples collected with dedicated triggers are used to extend the measurement of the jet energy spectrum to $E_T^{jet}\sim$100~GeV. 
Results for the ratio $\cal{R}$ obtained from an analysis of these samples similar to that described above are presented in Fig.~\ref{fig:sf_q2} for $Q^2\equiv \langle E_T^*\rangle^2$ in the range of $\sim 100<Q^2<10^4$~GeV$^2$. 
These distributions are mainly affected by the overall uncertainty in normalization, as the relative uncertainties among different $Q^2$ bins  cancel out in measuring the ratio. A factor of $\lesssim 2$ variation among all distributions is observed over the entire $Q^2$ range, as compared to a factor of $\sim 10^4$ over the same range between the individual SD and ND distributions shown in Fig.~\ref{fig:JetEtAve}. The results of fits performed using the form ${\cal R} = {\cal R_{\rm 0}}\cdot x_{Bj}^{r}$ in the region $1\times 10^{-3}<x_{Bj}<2.5\times 10^{-2}$, where the constraint $\beta=(x/\xi)<0.5$ is satisfied, are presented for various jet $E_T^*$ intervals in Table~\ref{tab:q2_summary}. Both ${\cal{R}}_0$ and $r$ are constant within the uncertainties over the range of $10^2<Q^2<10^4$ GeV$^2$. This result indicates that the $Q^2$ evolution in diffractive interactions is similar to that in ND interactions~\cite{blois07_cdf}.

\begin{table}[htb]
\begin{center}
\caption{\label{tab:q2_summary} Fit parameters of the ratio of SD to ND production rates for events in different $E_T^*$ bins. The ratios are fitted to the form ${\cal R} = {\cal R}_0\cdot x^r_{Bj}$ for $1\times 10^{-3}<x_{Bj}<2.5\times 10^{-2}$.}
\vspace*{0.5em}
\begin{tabular}{cccc}\hline\hline
Jet $E_T^*$ [GeV]	& $Q^2$ GeV$^2$	& ${\cal R}_0$ 	& $r$ \\ \hline
 $8<E_T^*<12$ 	& 100			& $(8.6\pm 0.8)\times10^{-3}$	& $-0.44\pm0.04$ \\ 
$18<E_T^*<25$ 	& 400			& $(8.0\pm 1.6)\times10^{-3}$	& $-0.48\pm0.05$ \\ 
$35<E_T^*<50$ 	& 1,600			& $(6.3\pm 1.8)\times10^{-3}$	& $-0.60\pm0.07$ \\ 
$50<E_T^*<70$ 	& 3,000			& $(5.5\pm 5.0)\times10^{-3}$	& $-0.64\pm0.22$ \\ 
$70<E_T^*<90$ 	& 6,000			& $(7.0\pm 7.0)\times10^{-3}$	& $-0.58\pm0.26$ \\ 
\hline\hline
\end{tabular}
\end{center}
\end{table}
 
\subsection{$t$ distribution\label{tdist}}
The $t$ distribution is measured for both soft and hard SD data samples using the inclusive RPS$_{\rm track}$ and the RPS$_{\rm track}\cdot$Jet5 (Jet20, Jet50) data.
In Sec.~\ref{tdist_1}, we discuss the $t$ distribution in the region of $0<|t|<1$~GeV$^2$ as a function of jet $\langle E_T^*\rangle^2\approx Q^2$ and characterize the slope parameter at $|t|\approx 0$ as a function of $Q^2$. Then, in Sec.~\ref{tdist_2}, we examine the shape of the $t$ distribution in the region up to  $|t|=4$~GeV$^2$ and search for a diffraction minimum. 
Crucial for the measurement at $|t|>4$ GeV$^2$ is the background subtraction.

\paragraph{Background evaluation.} The background is evaluated from the data by taking advantage of the asymmetric vertical position of the RPS detectors during data taking. As shown in Fig.~\ref{fig:offset}, the dynamic alignment method yields Y$_{\rm offset}\approx 0.25$~cm and X$_{\rm offset}\approx 0$. The nonzero value of Y$_{\rm offset}$ results in an asymmetric $t$ range in the acceptance. Since the RPS detectors are 2~cm wide , and given that $t\propto -\theta_{\bar p} ^2$, where $\theta_{\bar p} \propto$Y$_{\rm offset}$, the $|t|$-range of the RPS acceptance is expected to be a factor of $[(1+0.25)/(1-0.25)]^2=2.8$ larger in the  Y$_{\rm track}>{\rm Y}_0$ than in the Y$_{\rm track}<{\rm Y}_0$ region, where Y$_0$ is the Y$=0$ horizontal centerline of the detector.

\begin{figure}[htp]
\begin{center}
\vspace*{-0.5em}
\includegraphics[width=0.49\textwidth]{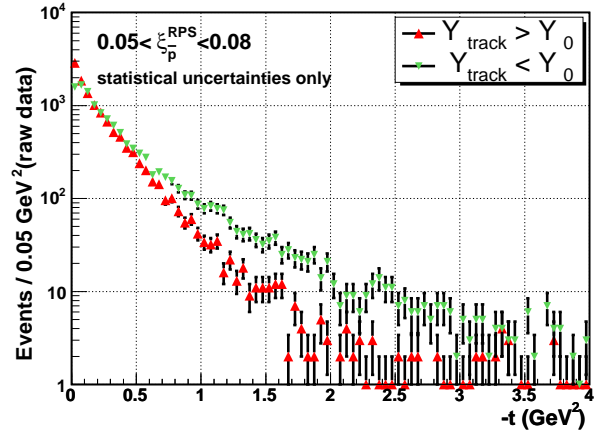}
\caption{
$t$ distributions of SD RPS$_{\rm track}$ dijet events within $0.05<\xi_{\bar{p}}^{\rm RPS}<0.08$ for Y$_{\rm track}>$Y$_0$ (up-pointing-triangles) and Y$_{track}<$Y$_0$ (down-pointing-triangles), where Y$_0$ is the centerline of the RPS fiber tracker.
}
\label{fig:tdistr_asymm} 
\end{center}
\end{figure}

Figure~\ref{fig:tdistr_asymm} shows $t$ distributions of 
SD RPS$_{\rm track}$ data within $0.05<\xi^{\rm RPS}_{\bar p}<0.08$. Two distributions are shown, one for Y$_{\rm track}>{\rm Y}_0$ and another for Y$_{\rm track}<{\rm Y}_0$. The Y$_{\rm track}>{\rm Y}_0$ distribution falls nearly exponentially from $t=0$ down to a value $|t_1|\approx 2.3$~GeV$^2$ and becomes consistent with an average flat background of $N_{\rm bg}\approx 20/{\rm GeV}^2$ in the unphysical region of $|t|>|t_1|$, in which the reconstructed track is outside the RPS acceptance. We have verified that the event rate at $|t|>t_1$ scales with that at $t\sim 0$ for runs of various instantaneous luminosities, and therefore conclude that these events represent a background associated with a $\bar p p $ interaction. However, since the reconstructed track is outside the RPS acceptance, its origin appears to be a particle from the $\bar pp$ interaction that suffered a secondary collision before reaching the RPS and produced a particle that traversed the RPS. Such secondary collisions would be expected to produce a flat distribution in the vicinity of the RPS in a plane perpendicular to the beam, consistent with the flat distribution observed at $|t|>2.5\;{\rm GeV}^2$ for Y$_{\rm track}>{\rm Y}_0$.

Based on the estimated value of $|t_1|\approx 2.3$~GeV$^2$ and the scaling factor of 2.8 for the detector acceptance due to the vertical misalignment, the unphysical region for the Y$_{\rm track}<{\rm Y_0}$ data is expected to be at $|t|>|t_2|$, where $|t_2|=2.3\times 2.8=6.5$~GeV$^2$. 
However, in order to reduce systematic uncertainties on acceptance corrections arising from ($t,\xi$)-resolution and detector edge effects we present in Sec.~\ref{tdist_2} $|t|$ distributions only up to $|t|=4$~GeV$^2$.

\begin{figure*}
\begin{center}
\hbox{\includegraphics[width=0.5\textwidth]{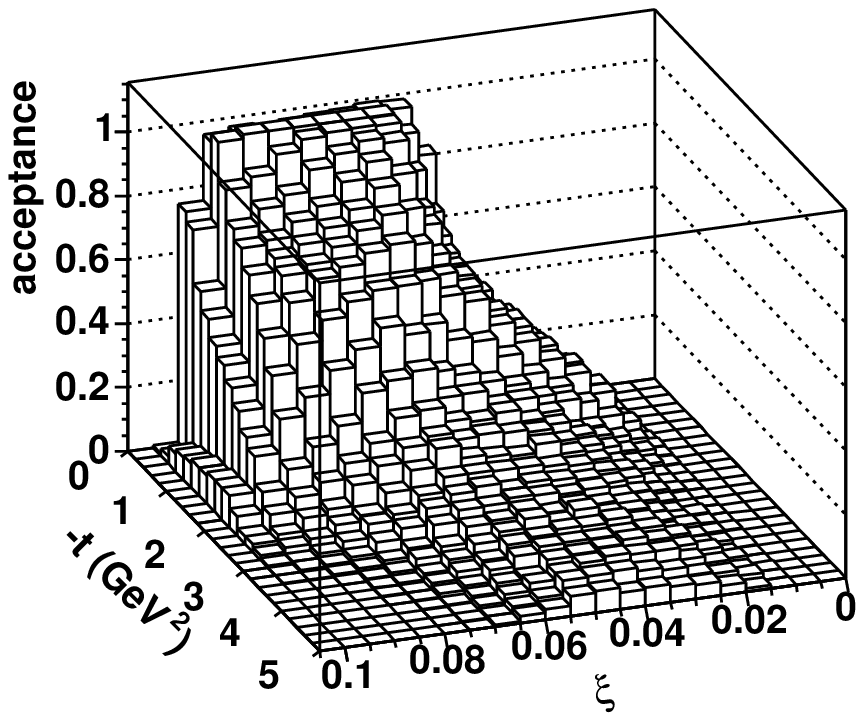},\includegraphics[width=0.45\textwidth]{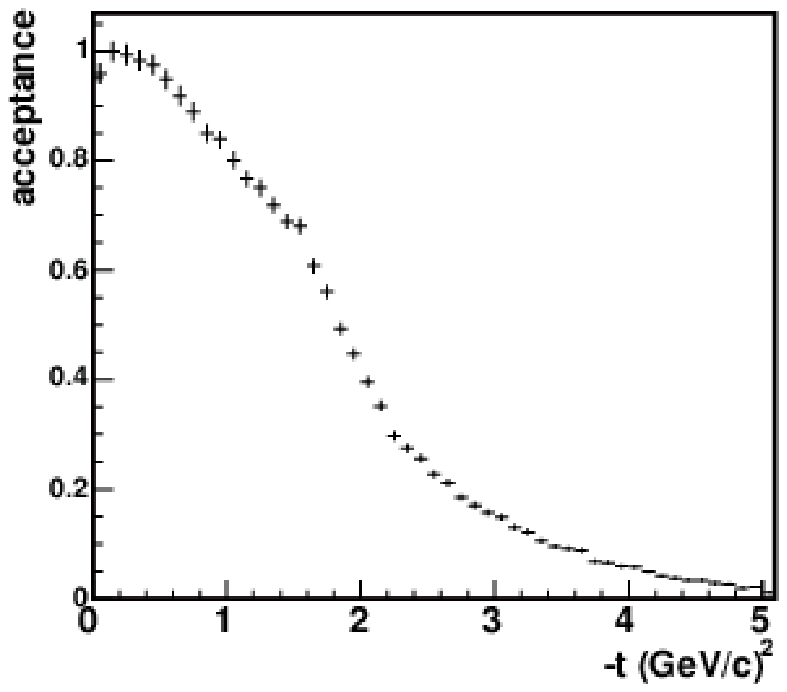}}
\caption{RPS acceptance: (left) vs $-t$ and $\xi$, and (right) vs $-t$ integrated over $0.05<\xi<0.08$.}
\label{fig:RPSacceptance}
\end{center}
\end{figure*}

\paragraph{Corrections.} The data are corrected by subtracting the background and dividing by the RPS acceptance. In principle, the background subtraction should be performed on an event-by-event basis, since both background and acceptance depend on both $\xi$ and $t$. The acceptance, $A(\xi,t)$, is obtained from a beam-optics simulation. 
For the data in the plateau region of the $\xi_{\bar p} ^{\rm RPS}$ distribution of Fig.~\ref{fig:xi_rp} within
$0.05<\xi<0.08$, which are used for evaluating $t$ distributions, the acceptance $A_{0.05<\xi<0.08}(t)$ is shown in Fig.~\ref{fig:RPSacceptance}. For each event, the raw values of $\xi$ and $t$ are obtained from the RPS tracker and used to fill the bins of a histogram. Since the acceptance is fairly flat versus  $\xi$ within the $t$ region of interest, the incremental value entered into a $t$-bin is reduced by the average background of $N_{bg}\times\Delta t$ events, where $\Delta t$ is the bin width, and is increased by a factor of $A^{-1}_{0.05<\xi<0.08}(t)$ (a zero value is entered if the number of events in a bin after subtracting the background is found to be $\leq 0$).

\subsubsection{$t$ distribution for $|t|\leq 1$ GeV$^2$}\label{tdist_1}
In this section, we discuss the event selection requirements used for the measurement of the $t$ distribution in the region of $|t|\leq 1$ GeV$^2$, extract the slope parameter(s) from fits to the data, and comment on systematic uncertainties.

\begin{figure}
\vspace*{-2em
\begin{center}}
\includegraphics[width=0.5\textwidth]{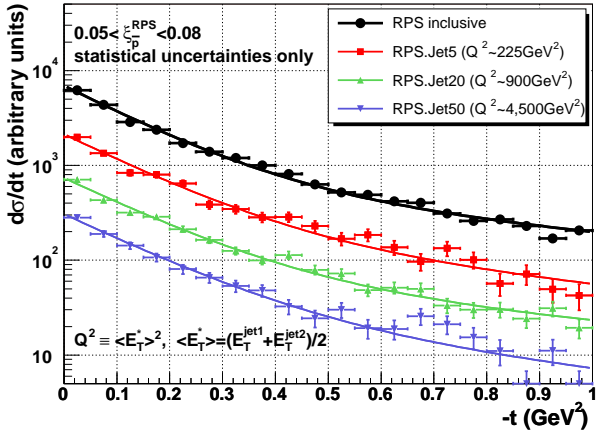}
\caption{\label{fig:bslope}$|t|_{\bar{p}}$ distribution shapes for SD RPS$_{\rm track}$ data of different average $Q^2$ values within $0.05<\xi_{\bar{p}}^{\rm RPS}<0.08$ (note the different $\xi^{\rm RPS}_{\bar{p}}$ range from that in Table~\ref{tab:SD_cuts}.}
\end{center}
\end{figure}

\paragraph*{\bf Event selection.} 
To minimize the effect of migration of events to and from adjacent $\xi$ bins caused by resolution effects, events are selected in the region of $|t|\leq 1$ GeV$^2$ and $0.05<\xi_{\bar p} ^{\rm RPS}<0.08$, where the RPS acceptance is approximately flat (see Fig.~\ref{fig:xi_rp}). To reject overlap backgrounds, these events are further required to be within $\xi_{\bar p} ^{\rm CAL}<0.1$, where the SD events dominate (see Fig.~\ref{fig:Xi_withRPtrack}).
The  same expression that was used in the RPS dynamic alignment in Sec.~\ref{sec:align}, composed of two exponential terms with slopes $b_1$ and $b_2$ (Eq.~\ref{eq:tslope_fit}), is used to fit the data and obtain the values of the slopes. Fits to data with only statistical uncertainties are shown in Fig.~\ref{fig:bslope} for various $Q^2$ ranges. Systematic uncertainties are discussed below.

\begin{figure}
\begin{center}
\hspace*{-2em}\includegraphics[width=0.51\textwidth]{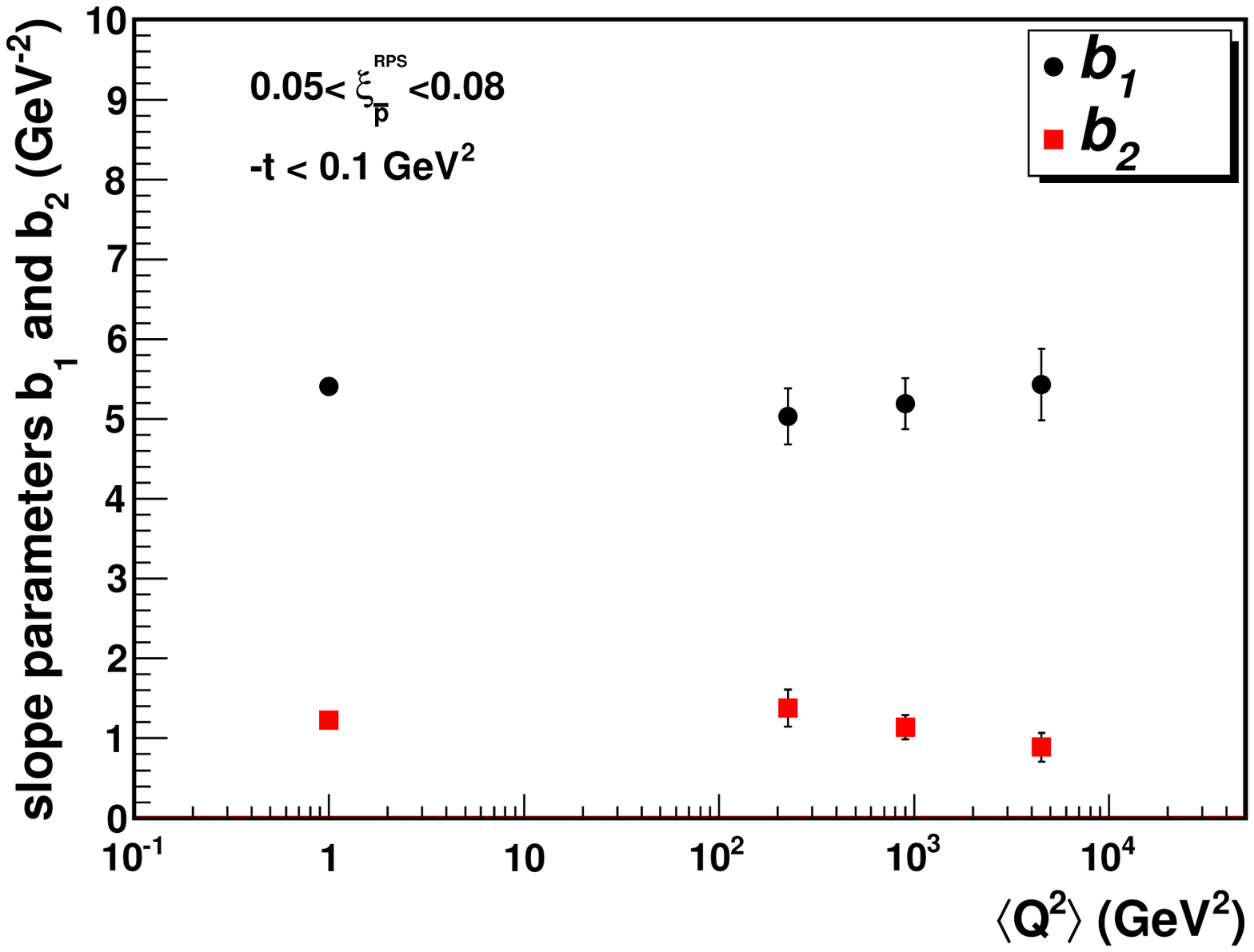}
\caption{The slope parameters $b_1$ and $b_2$ of a fit to the form $d\sigma/dt=N\cdot (A_1\cdot e^{b_1\cdot t}+A_2\cdot e^{b_2\cdot t})$, with $A_2/A_1=0.11$ for SD events of different $Q^2$ values (see Table~\ref{tab:q2_dep}); the soft diffraction (RPS inclusive) points have been placed arbitrarily at $<Q^2>$=1 GeV$^2$.}
\label{fig:bslopeQ2}
\end{center}
\end{figure}

\paragraph*{\bf \boldmath$b$-slope parameters.} Results for the slope parameters $b_1$ and $b_2$ and for the ratios $b_1/b_1^{\rm incl}$ and $b_2/b_2^{\rm incl}$ are presented in Table~\ref{tab:q2_dep}. The slopes $b_1$ and $b_2$ for all samples are plotted in Fig.~\ref{fig:bslopeQ2}.
No significant $Q^2$ dependence is observed from soft diffraction of $<Q^2>\sim 1$~GeV$^2$ to $Q^2\sim10^4$~GeV$^2$. 
The mean values of $b_1$  and $b_2$ over all data samples are $5.27\pm 0.33$ (GeV$/c)^{-2}$ and $1.17\pm 0.17$ (GeV$/c)^{-2}$, respectively. The measured slopes  of the inclusive sample are in agreement with theoretical expectations~\cite{ref:DL}. 

\begin{table*}[htb]
\begin{center}
\caption{\label{tab:q2_dep} Slopes of $t$ distributions for soft and hard diffractive events of the SD RPS$_{\rm track}$ data in the range $0.05<\xi_{\bar p}^{\rm RPS}<0.08$ for different $\langle E_T^*\rangle$ or $Q^2\equiv\langle E_T^*\rangle^2$ bins obtained from fits to the form of Eq.~(\ref{eq:tslope_fit}), $d\sigma/dt=N\cdot (A_1\cdot e^{b_1\cdot t}+A_2\cdot e^{b_2\cdot t})$, with $A_2/A_1=0.11$, fixed at the average value obtained in the dynamic alignment of all different event subsamples. The uncertainties listed are statistical.}
\vspace*{0.5em}
\begin{tabular}{lcccccc}\hline\hline
Event sample& $\langle E_T^*\rangle$ & $Q^2$  & $b_1$  & $b_2$& $b_1/$ $b_1^{\rm incl}$ 	& $b_2/$ $b_2^{\rm incl}$ \\ 
(definition)& (GeV) &  (GeV$^2$) & (GeV$^{-2}$) &(GeV$^{-2}$) &(ratio)&(ratio)\\ \hline
RPS 	& incl 	& $\approx 1$ &$5.4\pm 0.1$ 	& $1.2\pm 0.1$ & 1& 1\\
RPS$\cdot$Jet5			& 15 		& 225& $5.0\pm 0.3$	& $1.4\pm 0.2$&$0.93\pm 0.08$& $1.12\pm 0.23$\\
RPS$\cdot$Jet20			& 30 		& 900& $5.2\pm 0.3$ 	& $1.1\pm 0.1$&$0.96\pm 0.07$& $0.93\pm 0.16$\\
RPS$\cdot$Jet50			& 67 		& 4500 &$5.5\pm 0.5$ 	& $0.9\pm 0.2$&$1.00\pm 0.10$& $0.72\pm 0.18$\\
\hline\hline
\end{tabular}
\end{center}
\end{table*}

\paragraph*{\bf Systematic uncertainties.} 
We considered the dependence of the results on the RPS fiber-tracker thresholds, the instantaneous luminosity, and the beam-store and run number.

\setcounter{paragraph}{0}
\paragraph{RPS fiber-tracker thresholds.} Tracks in the RPS are reconstructed from hits above fiber threshold. 
A threshold value of 30 ADC counts is used for all three RPS stations. The slope values change by $+1\%$ ($-1\%$)  when the threshold is lowered down to 20 (raised up to 35) ADC counts, representing a decrease (increase) that doubles (reduces by 50\%) the number of noise hits in the fiber tracker. A systematic uncertainty of $\pm 1\%$ was assigned for this effect.  

\paragraph{Instantaneous luminosity.}
The RPS inclusive sample is divided into two luminosity-level subsamples: 
one consisting of events collected at low instantaneous luminosities, ${\cal L}< 1.5\cdot 10^{31}$cm$^{-2}$s$^{-1}$, 
and the other of events collected at ${\cal L}> 3.5\cdot 10^{31}$cm$^{-2}$s$^{-1}$.
A less than 2\% difference between the slopes extracted from the two samples is observed, which is well within the statistical uncertainty.

\paragraph{Beam conditions.}
Variations in beam conditions among various beam stores and/or during various time intervals of a given store may affect the slope measurement.
To estimate the uncertainty associated with varying beam conditions, we measure the $b$-slopes of data subsamples from several beam stores and from different periods (run number) within a store. The values for $b_1$ and $b_2$ obtained were within the corresponding statistical uncertainties. A systematic uncertainty equal to the statistical uncertainty was estimated.

\paragraph{Dynamic alignment.} A systematic uncertainty of $5\%$ is estimated from the fits shown in Fig.~\ref{fig:offset} for the slope $b_1$ arising from the uncertainties of $\pm 60\;\mu$m in the (X,Y) position of the RPS detectors relative to the $\bar p $ beam.

The systematic uncertainties are listed in Table~\ref{tab:tslope_syst}.

\begin{table}[htb]
\begin{center}
\caption{\label{tab:tslope_syst} Systematic uncertainties in the slope parameters $b_1$ and $b_2$ of the diffractive $t$-distributions (from Table~\ref{tab:q2_dep}).}
\begin{tabular}{l|c|r}\hline\hline
Source of uncertainty				&  $\delta b_1$&	$\delta b_2$\\ 
\hline
RPS tracker threshold				&  1\%	& 1\%  \\ 
Instantaneous luminosity			&  2\%	& 2\%  \\ 
Beam conditions				&  4\%	& 8\%  \\
RPS alignment					&  5\%  & 5\%  \\

Total (quadrature\,-\,sum)                          & 6.8\% &9.7\% \\
\hline\hline
\end{tabular}
\end{center}
\end{table}


\subsubsection{$t$ distributions for $|t|\leq4$ GeV$^2$ and  search for a diffraction minimum}\label{tdist_2}\hfill
We extend the analysis to measure $t$ distributions up to $|t|=4$~GeV$^2$ and search for a diffraction minimum using inclusive SD RPS$_{\rm track}$ and RPS$_{\rm track}\cdot$Jet5 (Jet20) events within $0.05<\xi_{\bar p}^{\rm RPS}<0.08$. This $\xi$-range corresponds to a mean mass for the diffractively dissociated proton $\langle M_X\rangle> \approx \xi\sqrt s\approx 500$~GeV.  
The dijets in the Jet5 (Jet20) data sample have $\langle Q^2\rangle\approx 225\;(900)\;{\rm GeV}^2$. As already discussed in Sec.~\ref{tdist}, the background in these data samples is $N_{bg}=20/{\rm GeV}^2$, as estimated from the unphysical region of $|t|>|t_1|=2.5\;{\rm GeV}^2$ for ${\rm Y}_{\rm track}>{\rm Y}_0$ in Fig.~\ref{fig:tdistr_asymm}. 

\begin{figure}[htp]
\begin{center}
\includegraphics[width=0.55\textwidth]{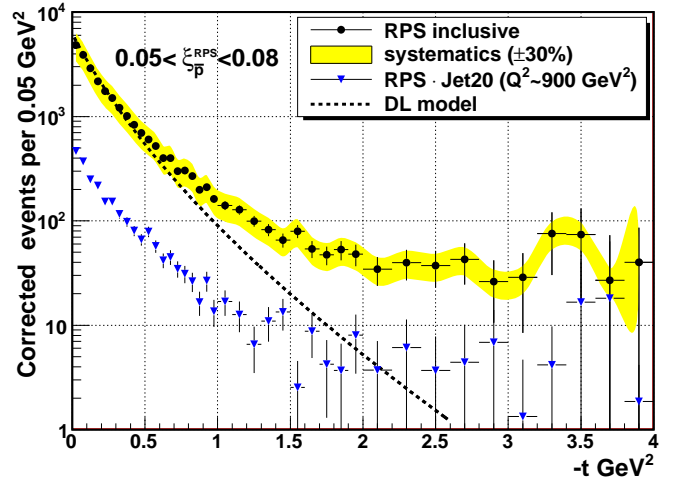}
\caption{$t$-distributions for two samples of SD RPS$_{\rm track}$ events within the region $0.05<\xi_{\overline{p}}^{\rm RPS}<0.08$ corrected for RPS acceptance after background subtraction: {(circles)} RPS inclusive and {(triangles)} RPS$\cdot$Jet20 ($\left<Q^2\right>\simeq900\;{\rm GeV}^2$). The curve represents the distribution expected for soft SD in the DL (Donnachie-Landshoff) model~\cite{ref:DL} (Eq.~\ref{eq:DL}) normalized to the RPS data within $-t\lesssim 0.5$~GeV$^2$.}
\label{fig:tdistr_hilum} 
\end{center}
\end{figure}

Figure~\ref{fig:tdistr_hilum} shows the sum of the Y$_{\rm track}>{\rm Y_0}$ and Y$_{\rm track}<{\rm Y_0}$ distributions for the RPS inclusive and RPS$\cdot$Jet20 data samples after background subtraction and acceptance corrections. The distributions are presented in a variable-bin-width-histogram format. In incrementing the histograms, a fraction is subtracted from each entry equal to the average background fraction in that bin, and the acceptance correction is then applied based on the RPS-measured values of $\xi$ and $t$. The dashed curve shown in the figure represents the electromagnetic form factor squared of the Donnachie-Landshoff (DL) model~\cite{ref:DL}, $F_1(t)^2$, normalized to the RPS inclusive data within $-t\lesssim 0.5$~GeV$^2$.
The $F_1(t)$ form factor used is given by~\cite{ref:DL}:

\begin{equation}
\label{eq:DLF1}
 F_1(t) = \frac{4m_p^2 - 2.8 t}{4 m_p^2 - t}       \left[ \frac{1}{(1-t/0.71)} \right]^2,
\end{equation}

\noindent In SD, $F_1(t)$  enters in the form~\cite{ref:DL}:

\begin{equation}
\label{eq:DL}
\frac{d\sigma^{SD}}{dt} = N_{norm}       F_1(t)^2       \exp\left[2\alpha' \cdot \ln\frac{1}{\xi}\cdot t\right],
\end{equation}

\noindent where $\alpha'\approx 0.25$~GeV$^{-2}$ is the slope of the $\Pomeron$-trajectory~\cite{Donnachie}. For $|t|\leq 1$~(GeV/c)$^2$ and $0.05<\xi<0.08$, the two-component exponential form of Eq.~(\ref{eq:tslope_fit}) with $A_2/A_1=0.11$, the average value obtained in the dynamic alignment method, is a good approximation to that of Eq.~(\ref{eq:DL}) with $\langle\xi\rangle=0.065$ substituted for $\xi$. 

The following features of the $t$ distributions are notable:

\begin{itemize}
\item {\em Low-$t$ region (\,$-t\lesssim 0.5$~{\rm GeV}$^2$):} the RPS data are in good agreement with the DL curve;
\item {\em Scale independence:} the distributions of the RPS and RPS$\cdot$Jet5 data are similar in shape;
\item {\em High-$t$ region (\,$-t\gtrsim0.5$~{\rm GeV}$^2$):} the RPS data lie increasingly higher than the DL curve as $-t$ increases, becoming approximately flat for  $-t\gtrsim 2$~GeV$^2$. The compatibility of this observation  with an underlying diffractive minimum at $-t\sim 2.5$~GeV$^2$ broadened by resolution effects is discussed below. 
\end{itemize}

The physics significance of these results is briefly discussed below.

\paragraph{Low-$t$ region.} The good agreement between the inclusive $t$ distribution and the DL prediction in this region serves as a basis for a search for deviations in the region of $|t|\gtrsim 0.5$~GeV$^2$ that could arise from a diffraction minimum.      

\paragraph{Scale independence.} The scale independence of the distributions supports a factorization property between the exchange that produces the leading $\bar{p}$ and associated rapidity gap on the one hand, and the final state into which the proton dissociates on the other. Such behavior favors models in which the hard scattering is controlled by the low-$x$ parton distribution function of the recoiling antiproton, just as in ND interactions, while a color-neutral soft exchange allows the antiproton to escape intact forming the rapidity gap (see, for example, Refs.~\cite{Goulianos}-\cite{Kopeliovitch}).     

\paragraph{High-{\em t} region.} In $\bar pp$ and $pp$ elastic scattering at collider energies, a diffraction minimum (dip) in the $t$ distribution is observed, with its value decreasing as $\sqrt s$ increases~(see, e.g., Ref.~\cite{Islam}). 
Recently, the D0 collaboration reported a preliminary Tevatron Run~II result on elastic $\bar{p}p$ scattering at $\sqrt s=1960$ GeV,  in which a dip (broadened by resolution effects) is observed at $-t\sim 0.7$ GeV$^2$ followed by a maximum (``bump'') at $-t\sim 1$ GeV$^2$~\cite{ref:D0elastic}. 
A dip in the $t$ distribution of $pp$ or $\bar pp$ diffraction dissociation has never been reported. In this analysis, since the quasielastic diffractive scattering occurs at $s'=\xi\cdot s<s$, the dip, if it exists, would be expected to lie at a higher $|t|$ than in elastic scattering and have a Gaussian-like width due to $\Delta\xi$-bin-size and $t$-resolution effects. 

The expected contributions to the width of a diffractive dip are summarized below:    
\begin{enumerate}[(a)]
\item $\xi$-bin width: from Eq.~(\ref{eq:DL}), $\Delta |t|_{\xi-{\rm bin}}=\Delta\ln(1/\xi)=\ln(1/0.05)-\ln(1/0.08)= 0.47$;
\item $\delta\xi$ and $\delta t$ resolutions: from Refs.~\cite{jjRPS,jjRPS630}, $\delta\xi=0.001$ and $\delta t=\pm 0.07$~GeV$^2$ for $\left<|t|\right>\approx 0.05$~GeV$^2$ with a dependence $\propto\sqrt{|t|}$, resulting for $|t|\approx 2.5$~GeV$^2$ in a value larger by a factor $\sqrt{2.5}/\sqrt{0.05}\approx{7}$ that yields $\delta t_{\rm res}\approx 0.5$~GeV$^2$; 
\item total width expected : $\Delta t_{|t|=2.5}=\Delta |t|_{\xi-{\rm bin}}+\delta t_{\rm res}\approx 1$~GeV$^2$, where the contributions have been added linearly, since the $\xi$-bin width is not a random variable.
\end{enumerate}

These estimates show that the rather flat $-t$ distributions at large $-t$ shown in Fig.~\ref{fig:tdistr_hilum} are compatible with a possible existence of an underlying diffraction minimum around $-t\sim 2.5$~GeV$^2$.

\section{Conclusion}\label{conclusion}
We present results from a measurement of diffractive dijet production in $\bar{p}p$ collisions at $\sqrt s=$~1.96~TeV, $\bar p+p\rightarrow [{\rm jet1}+{\rm jet2}+X]_{\bar p}+p$, extracted from data collected in Run~II of the Fermilab Tevatron $\bar{p}p$ Collider using the CDF~II detector. We also present a measurement of the $t$ distribution for single-diffractive events from an inclusive single diffractive data-sample up to $-t$=4 GeV$^2$, compare it to that of single diffractive dijet production, and search for a diffractive dip.
  
A system of special forward detectors including a Roman Pot Spectrometer (RPS) for measuring the momentum of the outgoing $\bar p$ were used. The measured diffractive dijet production rates as a function of  $x^{\overline{p}}_{Bj}$ ($x^{\overline{p}}$-Bjorken), $Q^2\approx \langle E_T\rangle^2$, $\xi_{\bar p}$, and $t_{\bar p}$ are compared with the corresponding nondiffractive rates
at the same $x^{\overline{p}}_{Bj}$ and $Q^2$. 
The physics interest in these measurements is to elucidate the QCD nature of the diffractive exchange, traditionally referred to as {\em Pomeron} exchange.    

Our dijet results are extracted from data samples of diffractive and nondiffractive events corresponding to 310 pb$^{-1}$ of integrated luminosity. To reduce systematic uncertainties, the diffractive and nondiffractive data were collected simultaneously using the same calorimeter trigger to accept high-$E_T$ jets, and were similarly analyzed. The measurements cover the region of $0.03<\xi_{\bar p}<0.09$, $|t_{\bar p}|\leq 4$ GeV$^2$, $0.001<x^{\bar p}_{Bj}<0.09$ and  $10^2$~GeV$^2<Q^2<10^4$~GeV$^2$. 

The $E_T^{\rm jet}$ distribution shapes are similar for single diffractive and nondiffractive events, and 
the  $x^{\overline{p}}_{Bj}$ distribution of the ratio of single diffractive to nondiffractive production rates is relatively flat in the region of $\beta\equiv x/\xi\leq 0.5$, where $\beta$ is the fraction of the momentum loss carried by the parton in the Pomeron that participates in the interaction.
A fit of the form ${\cal R} = {\cal R_\circ}\cdot x_{Bj}^{r}$ in the region of $0.001<x^{\bar p}_{Bj}<0.025$, where $\beta<0.5$, yields $r=-0.44\pm0.04$ and ${\cal R_\circ}= (8.6\pm 0.8)\times10^{-3}$, consistent with the Run~I CDF result of $r=-0.45\pm0.02$ and ${\cal R_\circ}= (6.1\pm 0.1)\times10^{-3}$ within the normalization systematic uncertainty of $\pm 30\%$ common to the Run~I and Run~II data samples.

The $Q^2$ dependence of the ratio of single diffractive to nondiffractive rates is relatively weak, varying by less than a factor of $\sim 2$ over the measured $Q^2$ range within which the individual single diffractive and nondiffractive distributions vary by a factor of $\sim 10^4$.

The $t$ distributions for $-t<1$~GeV$^2$, for the inclusive and all dijet event samples,  can be fit with a sum of two exponential terms of  the form ${d\sigma}/{dt}=N\cdot \left(A_1 \cdot e^{b_1\cdot t} +  A_2 \cdot e^{b_2\cdot t}\right)$. The slope parameters $b_1$ and $b_2$ are found to be independent of $Q^2$ over the range of $\sim 1$~GeV$^2$(inclusive)$<Q^2<10^4$ GeV$^2$. The mean values of $b_1$  and $b_2$ are $5.27\pm 0.33$ GeV$^{-2}$ and $1.17\pm 0.17$ GeV$^{-2}$, respectively. Using these slopes, the above double-exponential fit is in good agreement with the prediction from the Donnachie-Landshoff (DL) model~\cite{ref:DL} for $-t\lessapprox 0.5$~GeV$^2$.   

The search for a diffraction minimum in the $t$ distribution is conducted using data with $0.05<\xi_{\bar p}^{\rm RPS}<0.08$ and $-t$ up to 4~GeV$^2$. Three event-samples are studied,  the inclusive sample and two dijet samples of $\left< Q^2\right> \approx 225$ and $\approx 900$~GeV$^2$, respectively. The distributions for all three samples at $-t\gtrsim 1.5$~GeV$^2$ display a flattening-out relative to the electromagnetic form factor used in the DL model, possibly due to a diffraction minimum at $-t\sim 2.5$~GeV$^2$ filled by $t$-resolution and $\xi$-dispersion effects. 
At $-t\sim 2$~GeV$^2$, the measured event-rate for the inclusive-event sample is a factor of $\sim 10$ larger than the DL-model prediction normalized to the data distribution in the region of  $-t\leq 0.5$~GeV$^2$.

The relatively flat $x^{\bar p}_{Bj}$ distribution and the small $Q^2$ dependence of the diffractive to nondiffractive ratios, combined with the $Q^2$ independence of the $t$ distributions, favor models of hard diffractive production in which the hard scattering is controlled by the parton-distribution-function of the recoil antiproton while the rapidity-gap formation is governed by a color-neutral soft exchange~\cite{Goulianos}-\cite{Kopeliovitch}.       

\begin{acknowledgments}
We thank the Fermilab staff and the technical staffs of the participating institutions for their vital contributions. This work was supported by the U.S. Department of Energy and National Science Foundation; the Italian Istituto Nazionale di Fisica Nucleare; the Ministry of Education, Culture, Sports, Science and Technology of Japan; the Natural Sciences and Engineering Research Council of Canada; the National Science Council of the Republic of China; the Swiss National Science Foundation; the A.P. Sloan Foundation; the Bundesministerium f\"ur Bildung und Forschung, Germany; the Korean World Class University Program, the National Research Foundation of Korea; the Science and Technology Facilities Council and the Royal Society, UK; the Russian Foundation for Basic Research; the Ministerio de Ciencia e Innovaci\'{o}n, and Programa Consolider-Ingenio 2010, Spain; the Slovak R\&D Agency; the Academy of Finland; and the Australian Research Council (ARC). 
\end{acknowledgments}

\begin{thebibliography}{0}
\bibitem{elastic}F. Abe et al. (CDF Collaboration), 
Phys. Rev. D {\bf 50}, 5518 (1994).

\bibitem{sd}F. Abe {et al.} (CDF Collaboration), 
Phys. Rev D {\bf 50}, D{\bf 50} 5535 (1994).
 
\bibitem{total} F. Abe et al. (CDF Collaboration), 
Phys. Rev. D{\bf 50}, 5550 (1994).

\bibitem{jgj1995}F. Abe et al., (CDF Collaboration), 
Phys. Rev. Lett. {\bf 74}, 855 (1995).

\bibitem{W}F. Abe et al. (CDF Collaboration), 
Phys. Rev. Lett. {\bf 78}, 2698 (1997).

\bibitem{jj}F. Abe et al. (CDF Collaboration), 
Phys. Rev. Lett. {\bf 79}, 2636 (1997).

\bibitem{jgj1800}F. Abe et al. (CDF Collaboration), 
Phys. Rev. Lett. {\bf 80}, 1156 (1998).

\bibitem{jgj630}F. Abe et al., (CDF Collaboration), 
Phys. Rev. Lett. {\bf 81}, 5278 (1998).

\bibitem{b-quark}T. Affolder et al. (CDF Collaboration), 
Phys. Rev. Lett. {\bf 84}, 232 (2000).

\bibitem{jjRPS} T. Affolder et al. (CDF Collaboration), 
Phys. Rev. Lett. {\bf 84}, 5043 (2000).

\bibitem{dpdJJ}T. Affolder et al. (CDF C0llaboration),
Phys. Rev. Lett. {\bf 85}, 4215 (2000).

\bibitem{dd} T. Affolder et al. (CDF Collaboration), 
Phys. Rev. Lett. {\bf 87}, 141802-(1-6) (2001).

\bibitem{jpsi}T. Affolder et al. (CDF Collaboration), 
Phys. Rev. Lett. {\bf 87}, 241802-(1-6) (2001).

\bibitem{jjRPS630} D. Acosta et al. (CDF Collaboration), 
Phys. Rev. Lett. {\bf 88}, 151802-(1-6) (2002).

\bibitem{sdd}D. Acosta et al. (CDF Collaboration), 
Phys. Rev. Lett. {\bf 91}, 011802-(1-6) (2003).

\bibitem{idpe}D. Acosta et al. (CDF Collaboration), 
Phys. Rev. Lett. {\bf 93}, 141601-(1-7) (2004).

\bibitem{CDFgammagamma}T.~Aaltonen et al. (CDF Collaboration), 
Phys. Rev. Lett. {\bf 99}, 242002 (2007).

\bibitem{excl2j}T.~Aaltonen et al. (CDF Collaboration), 
Phys. Rev. D {\bf 77}, 052004 (2008). 

\bibitem{diffWZ}T.~Aaltonen {\em et al.} (CDF Collaboration), 
Phys. Rev. D {\bf 82}, 112004 (2010).

\bibitem{CDFgammagamma2}T.~Aaltonen {\em et al.} (CDF Collaboration), 
Phys. Rev. Lett. {\bf 108} 081801 (2012).

\bibitem{rapidity}Rapidity, $y=\frac{1}{2}\ln\frac{E+p_L}{E-p_L}$, and pseudorapidity, $\eta=-\ln\tan\frac{\theta}{2}$, where $\theta$ is the polar angle of a particle with respect to the proton beam ($+\hat z$ direction), are used interchangeably for particles detected in the calorimeters, since in the kinematic range of interest in this analysis they are approximately equal.


\bibitem{Collins}P.~D.~B. Collins, {\em An Introduction to Regge Theory and High Energy Physics}, Cambridge University  Press, Cambridge, (1977).

\bibitem{Barone}V.~Barone and E.~Predazzi, {\em High-Energy Particle Diffraction}, Springer Press, Berlin, (2002).

\bibitem{Donnachie}S.~Donnachie, G.~Dosch, P.~Landshoff, and O.~Nachtmann, {\em Pomeron Physics and QCD}, Cambridge University Press, Cambridge, (2002). 
      
\bibitem{dino}K.~Goulianos, ``Hadronic diffraction: where do we stand?'', in {\em Les Rencontres de Physique de la Vallee d'Aoste: Results and Perspectives in Particle Physics, La Thuile, Italy, February 27 - March 6, 2004}, Frascati Physics Series, Special 34 Issue, edited by Mario Greco; arXiv:hep-ph/0407035.

\bibitem{jetcorrections}F. Abe {\it et al.} (CDF Collaboration), 
Phys. Rev. D {\bf 45}, 1448 (1992).

\bibitem{ken_thesis}K.~Hatakeyama, ``Measurement of the Diffractive Structure Function of the Antiproton in Proton-Antiproton Collisions at $\sqrt s=1800$ and 630 GeV,'' Ph.D. thesis, The Rockefeller University, 2003 (FERMILAB-THESIS-2003-18).

\bibitem{MP_prototype_test}K.~Goulianos and Stefano Lami, 
Nucl. Instrum. Methods A {\bf 430}, 34 (1999).

\bibitem{forward_proposal}
M.G.~Albrow {\it et al.} (CDF Collaboration), ``Further Studies in Hard Diffraction and Very Forward Physics,'' Fermilab-Proposal-0916, Oct 1999; \url{http://lss.fnal.gov/archive/test-proposal/0000/fermilab-proposal-0916.shtml}.

\bibitem{miniplug_nim} K. Goulianos, M.~Gallinaro, K. Hatakeyama, S. Lami, C. Mesropian, A. Solodsky,
Nucl. Instrum. Methods A {\bf 496}, 333 (2003).

\bibitem{michele_rio}M.~Gallinaro for the CDF Collaboration, ``CDF Forward Detectors and diffractive structure functions at the Fermilab Tevatron,'' Proceedings of the 9th Hadron Physics and 8th Relativistic Aspects of Nuclear Physics (HADRON-RANP 2004), Angra dos Reis, Rio de Janeiro, Brazil, March 28-April 3, 2004; FERMILAB-CONF-04-308-E, arXiv:hep-ph/0407255.

 \bibitem{CLC}  D. Acosta {\it et al.} (CDF Collaboration), Nucl. Instrum. Methods A {\bf 494}, 57 (2002).

\bibitem{CDF-I} F.~Abe {\it et al.}  (CDF Collaboration), Nucl. Instrum. Methods A {\bf 271}, 387 (1988);

\bibitem{CDF-II} D.~Acosta {\it et al.}  (CDF Collaboration), Phys.\ Rev.\ D {\bf 71}, 032001 (2005).

\bibitem{COT}A. Affolder {\it et al.}, Nucl. Instrum. Methods A {\bf 526}, 249 (2004).

\bibitem{SVX}A. Sill {\it et al.}, Nucl. Instrum. Methods A {\bf 447}, 1 (2000).

\bibitem{CEM} L. Balka {\it et al.}, Nucl. Instrum. Methods A {\bf 267}, 272 (1988).

\bibitem{CHA} S. Bertolucci {\it et al.}, Nucl. Instrum. Methods A {\bf 267}, 301 (1988).

\bibitem{PCAL} M. Albrow {\it et al.}, Nucl. Instrum. Methods A {\bf 480}, 524 (2002).

\bibitem{MidPoint} G.C. Blazey {\it et al.}, {\it ``Run II Jet Physics: Proceedings of the Run II QCD and Weak Boson Physics Workshop''}; arXiv:hep-ex/0005012.

\bibitem{missingET} Transverse energy is defined as $E_T=E\sin\theta$, and missing $E_T$ as $\met = |\overrightarrow{\met}|$ with  $\overrightarrow{\met}= -\sum_iE_T^i\hat{n_i}$, where $\hat{n_i}$ is a unit vector perpendicular to the beam axis and pointing at the $i^{th}$ calorimeter tower. The sum $E_T$ is defined by $\sum E_T=\sum_iE_T^i$. Both sums are over all calorimeter towers above the set thresholds. The missing $E_T$ significance is defined as $S_{\not E}\equiv{\not\!E_T}/\sqrt{\Sigma{E_T^2}}$.

\bibitem{blois07_cdf}K.~Goulianos, ``Diffraction at CDF,'' in Proc.  {\em 12\protect$^{th}$ International Conference on Elastic and Diffractive Scattering - Forward Physics and QCD (EDS07)}, DESY, Hamburg, Germany 21-25 May 2007; \url{http://www.desy.de/~eds07/proceedings.html}.


\bibitem{ref:DL}A.~Donnachie and P.~Landshoff, 
Phys. Lett. B {\bf 518} (2001) 63 (and references therein).

\bibitem{Goulianos}K.~Goulianos, ``Renormalized Diffractive Parton Densities,''  PoS (DIFF2006) 044 (and references therein), Presented at {\em Diffraction 06, International Workshop on Diffraction in High-Energy Physics}\, September 5-10, 2006, Adamantas, Milos island, Greece.

\bibitem{Brodsky} Stanley J. Brodsky, Rikard Enberg, Paul Hoyer, and Gunnar Ingelman, 
Phys. Rev. D {\bf 71}, 074020 (2005). 

\bibitem{Khoze}A.~B.~Kaidalov, V.~A.~Khoze, A.~D.~Martin, and M.~G.~Ryskin,
Eur.\ Phys.\ J.\  C {\bf 21} (2001) 521.

\bibitem{Kopeliovitch}B.Z. Kopeliovich, I.K. Potashnikova, I. Schmidt, and A.V. Tarasov, 
Phys. Rev. D {\bf 76}, 034019 (2007).

\bibitem{Islam}M. M. Islam, R .J . Luddy, A. V. Prokudin, 
Phys. Lett. B {\bf 605} (2005) 115-122.

\bibitem{ref:D0elastic}Andrew Brandt (for the D0 Collaboration), ``D0 Measurement of the elastic $p\bar p$  differential cross section for $0.25<|t|<1.2$ GeV$^2$ at $\sqrt s= 1.96$~TeV," presented at{\em  XVIII International Workshop on Deep-Inelastic Scattering and Related Subjects}, April 19 -23, 2010 Convitto della Calza, Firenze, Italy; Published in PoS DIS2010:059, 2010; Fermilab-conf-10-547.pdf
\end{thebibliography}

\end{document}